\documentclass{aa}
\usepackage{amsmath}
\usepackage{graphicx}
\usepackage{color}
\usepackage{ulem}

\usepackage{txfonts}
\usepackage{epstopdf} 
\usepackage{siunitx}
\usepackage{caption}
\usepackage{booktabs}  

\usepackage{hyperref}

\usepackage{array}
\newcolumntype{L}[1]{>{\raggedright\let\newline\\\arraybackslash\hspace{0pt}}m{#1}}
\newcolumntype{C}[1]{>{\centering\let\newline\\\arraybackslash\hspace{0pt}}m{#1}}
\newcolumntype{R}[1]{>{\raggedleft\let\newline\\\arraybackslash\hspace{0pt}}m{#1}}

\begin{document} 

\title{\textit{Gaia} Early Data Release 3}

\subtitle{Modelling and calibration of \textit{Gaia}'s point and line spread
functions}

\titlerunning{\textit{Gaia} Early Data Release 3 -- Modelling \textit{Gaia}'s
PSF and LSF}
\authorrunning{Rowell et al.}

\author{
N.~Rowell\inst{\ref{inst:uoe}}\fnmsep\thanks{Corresponding author: N.
Rowell\newline
e-mail: \href{mailto:nr@roe.ac.uk}{\tt nr@roe.ac.uk}}
\and M.~Davidson\inst{\ref{inst:uoe}}
\and L.~Lindegren\inst{\ref{inst:lund}}
\and F.~van Leeuwen\inst{\ref{inst:ioa}}
\and J.~Casta{\~n}eda\inst{\ref{inst:dapcom}}
\and C.~Fabricius\inst{\ref{inst:ieec}}
\and U.~Bastian\inst{\ref{inst:ari}}
\and N.~C.~Hambly\inst{\ref{inst:uoe}}
\and J.~Hern{\'a}ndez\inst{\ref{inst:esac}}
\and A.~Bombrun\inst{\ref{inst:hespaceesac}}
\and D.~W.~Evans\inst{\ref{inst:ioa}}
\and F.~De Angeli\inst{\ref{inst:ioa}}
\and M.~Riello\inst{\ref{inst:ioa}}
\and D.~Busonero\inst{\ref{inst:oato}}
\and C.~Crowley\inst{\ref{inst:hespaceesac}}
\and A.~Mora\inst{\ref{inst:auroraesac}} 
\and U.~Lammers\inst{\ref{inst:esac}}
\and G.~Gracia\inst{\ref{inst:poesac}\ref{inst:ari}}
\and J.~Portell\inst{\ref{inst:ieec}}
\and M.~Biermann\inst{\ref{inst:ari}}
\and A.~G.~A.~Brown\inst{\ref{inst:leiden}}
}  


\institute{
Institute for Astronomy, School of Physics and Astronomy, University of Edinburgh, 
Royal Observatory, Blackford Hill, Edinburgh, EH9~3HJ, 
United Kingdom
\label{inst:uoe} 
\and
Lund Observatory, Department of Astronomy and Theoretical Physics, Lund University, Box 43, SE-22100, Lund, Sweden
\label{inst:lund} 
\and
Institute of Astronomy, University of Cambridge, Madingley Road, Cambridge CB3~0HA, UK
\label{inst:ioa} 
\and
DAPCOM for Institut de Ciències del Cosmos (ICCUB), Universitat de Barcelona
(IEEC-UB), Martí Franquès 1, E-08028 Barcelona, Spain
\label{inst:dapcom} 
\and
Institut de Ci\`encies del Cosmos (ICCUB), Universitat de Barcelona (IEEC-UB),
Mart\'i Franqu\`es 1, E-08028 Barcelona, Spain 
\label{inst:ieec} 
\and
Astronomisches Rechen-Institut, Zentrum f\"ur Astronomie der Universit\"at Heidelberg, M\"onchhofstra{\ss}e 14, 69120 Heidelberg,
Germany
\label{inst:ari} 
\and
ESA, European Space Astronomy Centre, Camino Bajo del Castillo s/n, 28691
 Villanueva de la Ca{\~n}ada, Spain
\label{inst:esac} 
\and
HE Space Operations BV for ESA/ESAC, Camino Bajo del Castillo s/n, 28691 Villanueva de la Ca{\~n}ada, Spain
\label{inst:hespaceesac} 
\and
Gaia Project Office for DPAC/ESA, Camino Bajo del Castillo s/n, 28691 Villanueva de la Ca{\~n}ada, Spain
\label{inst:poesac} 
\and
Aurora Technology for ESA/ESAC, Camino Bajo del Castillo s/n, 28691 Villanueva de la Ca{\~n}ada, Spain
\label{inst:auroraesac} 
\and
Istituto Nazionale di Astrofisica, Osservatorio Astrofisico di Torino, Via Osservatorio 20, Pino Torinese, Torino, 10025, Italy
\label{inst:oato} 
\and
Leiden Observatory, Leiden University, Niels Bohrweg 2, 2333 CA Leiden, The
Netherlands
\label{inst:leiden} 
} 

\date{}

 

\abstract{
{\it Context} The unprecedented astrometric precision of the \textit{Gaia}
mission relies on accurate estimates of the locations of sources in the
\textit{Gaia} data stream. This is ultimately performed by point spread
function (PSF) fitting, which in turn requires an accurate reconstruction of the
PSF, including calibrations of all the major dependences. These include a strong
colour dependence due to \textit{Gaia}'s broad $G$ band and a strong time
dependence due to the evolving contamination levels and instrument focus.
Accurate PSF reconstruction is also important for photometry.
\textit{Gaia} Early Data Release 3 (EDR3) will, for the first time, use a PSF
calibration that models
several of the strongest dependences,
leading to signficantly reduced systematic errors.\\
{\it Aims} We describe the PSF model and calibration pipeline implemented for
\textit{Gaia} EDR3, including an analysis of the calibration results over the 34
months of data. We include a discussion of the limitations of the current pipeline and
directions for future releases. This will be of use both to users of
\textit{Gaia} data and as a reference for other precision astrometry missions.\\
{\it Methods} We develop models of the 1D line spread function (LSF) and 2D
PSF profiles based on a linear combination of basis components. These are designed for flexibility
and performance, as well as to meet several mathematical criteria such as
normalisation. We fit the models to selected primary sources in independent
time ranges, using simple parameterisations for the colour and other
dependences. Variation in time is smoothed by merging the independent
calibrations in a square root information filter, with resets at certain
mission events that induce a discontinuous change in the PSF.\\
{\it Results} The PSF calibration shows strong time and colour dependences that
accurately reproduce the varying state of the \textit{Gaia} astrometric
instrument.
Analysis of the residuals reveals both the performance and the limitations of
the current models and calibration pipeline,
and indicates the directions for future development.\\
{\it Conclusions} The PSF modelling and calibration carried out for
\textit{Gaia} EDR3 represents a major step forwards in the data processing and
will lead to reduced systematic errors in the core mission data products.
Further significant improvements are expected in the future data releases.
}


\maketitle

\section{Introduction}
\textit{Gaia} Early Data Release 3 (EDR3), the third release of data from the
European Space Agency mission \textit{Gaia} \citep{2016A&A...595A...1G}, contains
results based on data collected during the first 34 months of the nominal
mission
\citep{EDR3-DPACP-130}.
The principles of the \textit{Gaia} cyclic data processing are such that each
successive release is based on a complete reprocessing of the mission data collected
up to the chosen cutoff point.
This allows the released data to benefit from substantial improvements to
various core charge-coupled device (CCD) calibration and instrument models made
during the mission as the understanding of the payload develops. This is
crucial in beating down systematic errors present in earlier data releases and
leads to improvements in the astrometric and photometric data that are better
than expected based purely on the increased quantity of raw observations.
Of the core CCD calibrations---including, for example, the bias prescan
and non-uniformity
\citep{pemnu},
straylight and CCD health---the calibration and modelling of the point and
line spread functions (PSF and LSF) is perhaps the most vital in terms of
improving both the accuracy and precision of single-observation measurements of
the location and $G$ band flux of sources in the \textit{Gaia} data stream,
which are the quantities used to drive the astrometric and photometric
($G$ band) solutions.
We note that the LSF simply refers to the 1D image of a point source obtained by
marginalising a 2D image over one dimension, which is how the majority of
\textit{Gaia} observations are made. The PSF and LSF are modelled and calibrated
independently. Throughout this paper we use the acronym PLSF when referring to
both the point and line spread functions.

The central goal of the PLSF calibration is to produce a model of a given
stellar observation that can be used to estimate, via a separate process of
image parameter determination (IPD; see section~\ref{sec:ipd}), the
instrumental flux and location of the source in the \textit{Gaia} data stream.
The detailed shape of the PLSF varies significantly with time, colour, and
position in the focal plane and has numerous additional dependences of varying
significance, some of which are unique to \textit{Gaia}. The accurate
calibration of these effects in the PLSF is vital for the elimination of
systematic errors in both the astrometry and photometry.
In principle, the biases introduced by these effects can be reduced either in a
statistical way during the astrometric and photometric calibration, where they
manifest as structure in the residuals to the calibration, or in a direct way by
absorbing them into the PLSF calibration.
While the former approach was used in \textit{Gaia} Data Release 2 (DR2)
\citep[see e.g.][section 3.3]{dr2_astro}, the ultimate goal is to pursue the
latter approach, and EDR3 represents the first major step in achieving this.
This approach enables a clear demarcation of the roles of the PLSF calibration
and the astrometric and photometric calibrations, it provides the best PLSF
model for use in more sophisticated image analyses, for example of extended
objects or close binaries, and it allows the proper handling of less common
types of observation, such as those that use non-nominal instrument
configurations, for which the statistical approach does not correctly remove the
biases.

In this paper we present the modelling and calibration of \textit{Gaia}'s PSF
and LSF carried out in support of EDR3. We note that these results also apply to
the full DR3, as the calibrations will not be updated and the astrometry and
(integrated) photometry included in EDR3 (and which make use of the PLSF
calibrations) will not be recomputed.
The PLSF models presented here are not the only ones used in the
\textit{Gaia} data processing. An independent model is implemented within the
real-time processing systems devoted to internal scientific validation of the
astrometric processing chain \citep[see][Section 3.5.2]{EDR3-DOC}. The PLSF
models presented in this paper have been adopted in the cyclic data processing
for use in production of the data releases.
\section{Description of the instrument and observations \label{sec:instrument}}
In this section we briefly review the main properties of the \textit{Gaia}
optical system, CCDs and observations that are relevant to the PLSF modelling
and calibration. A more detailed description can be found in section 2 of
\citet{dr1_preprocessing}, to which the reader is referred for further
information.
\subsection{The optical system and focal plane instruments}
\begin{figure}
\centering
  \includegraphics[width=0.45\textwidth]{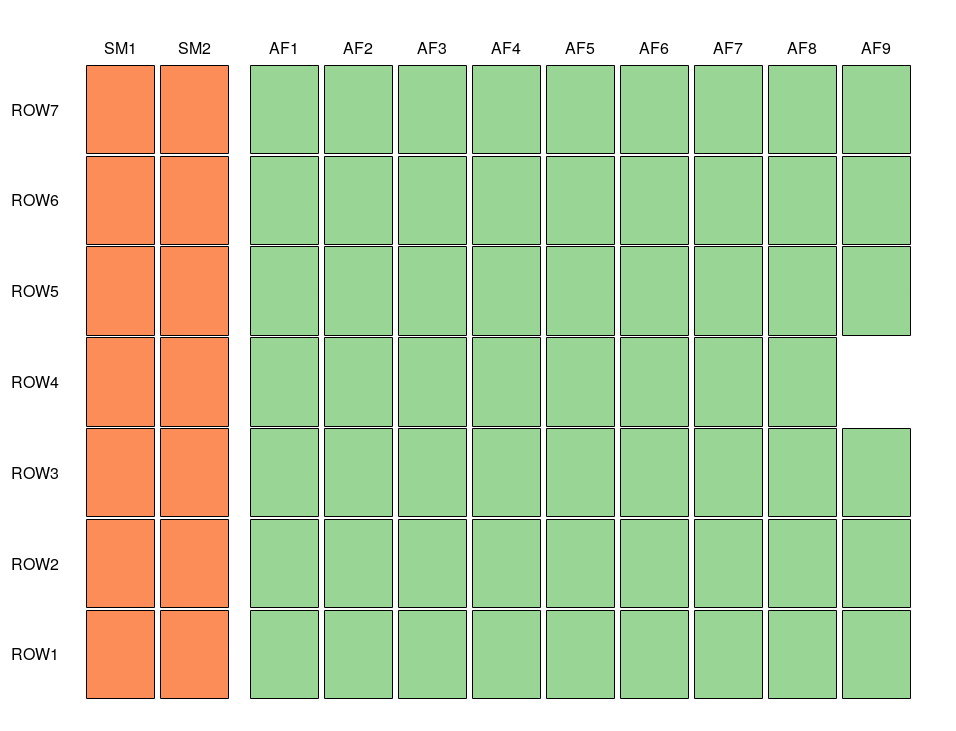}
  \caption{Schematic of the 14 sky mapper (orange) and 62 astrometric field
  (green) CCDs and their arrangement in the focal plane. In this diagram, stars
  enter the focal plane from the left and drift slowly to the right over around
  a minute.}
    \label{fig:smAf}
\end{figure}

The \textit{Gaia} instruments consist of two telescopes separated by a wide
`basic angle' of $106.5^{\circ}$ that form images on a single shared focal plane
array of 106 CCDs. The CCDs are arranged into seven rows containing 13--17
columns or `strips'; the CCDs in each row are divided among several instruments
that are used to perform measurements for scientific or diagnostic purposes.
In this paper we are concerned with modelling the LSF and PSF of two of the
instruments, the Sky Mapper (SM) and the Astrometric Field (AF), which both
observe unfiltered and undispersed light in the \textit{Gaia} $G$ band. The SM
and AF consist of 14 and 62 CCDs respectively. The arrangement of these CCDs in
the focal plane and their designation is depicted in Figure~\ref{fig:smAf}; the
CCDs are oriented such that the readout direction is to the right.
The satellite rotates about an axis perpendicular to the plane of the telescope
axes, with a rotation period of \textasciitilde6 hours. Stellar images drift
across the focal plane from left to right in Figure~\ref{fig:smAf} over a period
of around a minute. The directions in the focal plane parallel and perpendicular
to the stellar motion are referred to as the along-scan (AL) and across-scan
(AC) directions respectively.
The CCDs are operated in time-delayed integration (TDI) mode, where charge is
transferred slowly at a rate that matches the stellar motion, thus allowing the
charge to accumulate during readout.
The TDI period, corresponding to the time taken for the charge to be
transferred by one pixel in the AL direction, is $0.9828$
milliseconds. The onboard mission timeline (OBMT) is used to define
the timing of events on the satellite; for convenience, in this paper we express
OBMT in units of 6 hours, or one revolution (rev), which corresponds
(approximately) to the satellite rotation period\footnote{A tool for
transformation between OBMT and other time systems is available at
\texttt{https://gaia.esac.esa.int/decoder/obmtDecoder.jsp}; see also Equation 1
in \citet{dr2_astro}.
}.
%
%
The OBMT range covered by the EDR3 input data corresponds to 1078--5230 revs, or
\textasciitilde34 months.

\subsection{Observations of stars}
The two telescopes are referred to as field of view 1 (FOV1) and 2 (FOV2). Each
strip of SM CCDs sees the light from only one of the two FOVs (FOV1 for SM1 and
FOV2 for SM2), then the two FOVs are superimposed on the AF strips. Each source
that transits the focal plane thus has either nine (ROW4) or ten (ROW1--3,5--7)
$G$ band observations. The SM and AF1 CCDs are used in real-time to autonomously
detect the presence of stars and other astronomical objects, estimate their
magnitude and to predict their motion over the focal plane.
This enables the use of windowing to sample the data and downlink only small
sections surrounding each detected source. This optimises the available
telemetry budget, as well as reducing the readout noise, at the expense of
introducing some complications in the processing. In addition, for the great
majority of windows on-chip binning is used to marginalise the AC dimension so
that only a 1D profile in the AL direction is observed.
This is consistent with the astrometric observing principles of \textit{Gaia}
(and indeed of Hipparcos), for which an accurate location in the AL direction is
far more important.
Each observation is assigned a window with a particular geometry according to
its estimated magnitude and the CCD strip. The windowing is of fundamental
importance to the PLSF calibration and the specification of the windows by
magnitude and strip is shown in Table~\ref{tab:windows}. The different window
geometries split the data naturally into subsets that are labelled by
window class (WC); in SM there are two window classes labelled WC0 and
WC1, whereas in AF there are three window classes: WC0, WC1, and WC2. The pixel
scale is such that a window of 18$\times$12 pixels has an angular size of
roughly $1\farcs1\times2\farcs1$ AL$\times$AC.
\begin{table*}
\centering
\captionof{table}{Specification of the downlinked window geometry by onboard
estimated $G$ magnitude and CCD strip. In each case the window class (WC)
is given followed by the number of samples in the AL and AC directions, the
number of pixels in the AL and AC directions that have been binned to produce
each sample (in brackets), and whether the resulting observation
corresponds to the LSF or PSF. The solid
black lines indicate the geometry of each sample and their arrangement to form the
window; the faint dashed lines indicate the pixels that have been binned to
produce each sample. A combination of on-chip and numerical (software) binning
is applied to optimise onboard performance and telemetry budget.}
\label{tab:windows}
\begin{tabular}{|l|C{5cm}|C{4cm}|C{4cm}|}
	\toprule
	\hline
    Strip & $G$<13 & 13<$G$<16 & 16<$G$ \\ \hline
    SM &
    \begin{minipage}[c][2cm][c]{5cm}
    \centering
    WC0: 40$\times$6 (2$\times$2) PSF \newline
    \includegraphics[height=1.5cm]{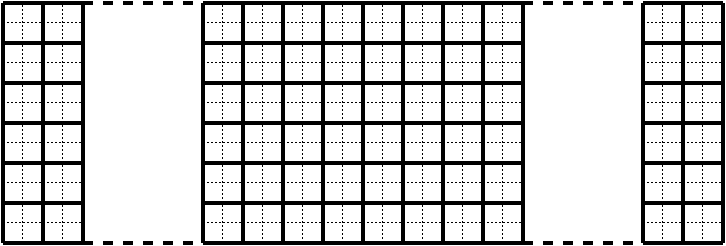}
    \end{minipage}
    &
    \multicolumn{2}{C{8cm}|}{
    \begin{minipage}[c][2cm][c]{4cm}
    \centering
    WC1: 20$\times$3 (4$\times$4) PSF \newline
    \includegraphics[height=1.5cm]{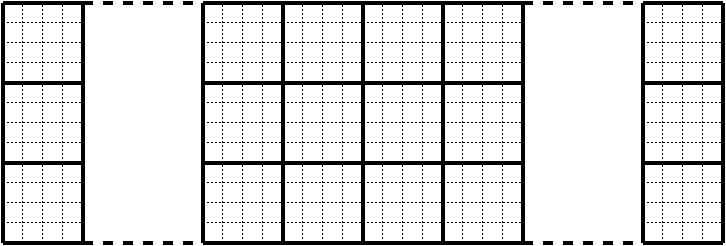}
    \end{minipage}
    }
    \\ \hline
    AF1 &
    \begin{minipage}[c][2cm][c]{5cm}
    \centering
    WC0: 18$\times$6 (1$\times$2) PSF \newline
    \includegraphics[height=1.5cm]{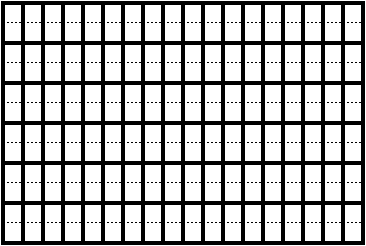}
    \end{minipage}
    &
    \begin{minipage}[c][2cm][c]{4cm}
    \centering
    WC1: 12$\times$1 (1$\times$12) LSF \newline
    \includegraphics[height=1.5cm]{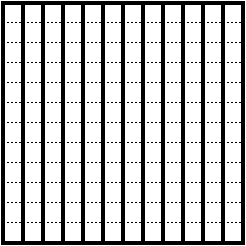}
    \end{minipage}
    &
    \begin{minipage}[c][2cm][c]{4cm}
    \centering
    WC2: 6$\times$1 (1$\times$12) LSF \newline
    \includegraphics[height=1.5cm]{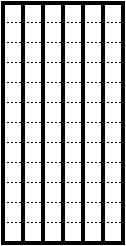}
    \end{minipage}
    \\ \hline
    AF2--9 &
    \begin{minipage}[c][2cm][c]{5cm}
    \centering
    WC0: 18$\times$12 (1$\times$1) PSF \newline
    \includegraphics[height=1.5cm]{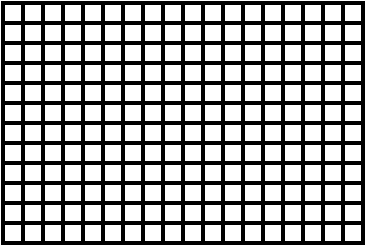}
    \end{minipage}
    &
    \begin{minipage}[c][2cm][c]{4cm}
    \centering
    WC1: 18$\times$1 (1$\times$12) LSF \newline
    \includegraphics[height=1.5cm]{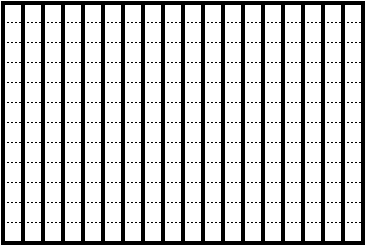}
    \end{minipage}
    &
    \begin{minipage}[c][2cm][c]{4cm}
    \centering
    WC2: 12$\times$1 (1$\times$12) LSF \newline
    \includegraphics[height=1.5cm]{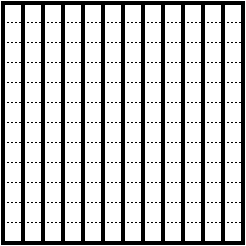}
    \end{minipage}
    \\ \hline \bottomrule
\end{tabular}
\end{table*}
The combination of 2D and 1D windows necessitates the calibration of both the
PSF, used to model the 2D observations, and the LSF, used to model the 1D
observations.

The \textasciitilde$4.42$ seconds that it takes a stellar image to cross an
individual CCD places an upper limit on the integration time available for each
source. For such an observation, the charge is accumulated over the entire AL
range of the CCD. In order to expand the dynamic range of \textit{Gaia}, bright
stars that are expected to saturate the detector can be observed for a shorter
integration time by the activation of special structures in the CCDs known as
TDI gates. These are positioned at a range of AL locations along the readout
direction and can be used to temporarily hold back the charge transfer, thus
resetting the charge accumulation and reducing the integration time. The gate
assignment and activation is done autonomously in real time based on the onboard
estimated magnitude from the SM observation.
There are eight gates routinely in use, referred to as gates 4, 7, 8, 9, 10, 11,
12, and 0 \citep[see][figure 5]{2016A&A...595A...1G}. 
The gating strategy maximises charge collection and minimises saturation for
bright stars spanning three orders of magnitude in brightness. The exposure time
varies from $0.016$ seconds for gate 4 up to $4.42$ seconds when no gate is
applied (gate 0). For gates 7 to 11, the exposure time is approximately doubled
for each transition.
Observations taken with each gate effectively sample a different AL range of the
CCD and are therefore expected to have a different PLSF, due to spatial
variations within the detector area. This further splits the data into subsets
that are labelled by TDI gate.
Only sources with $G\lesssim12.5$ are bright enough for TDI gates to be
activated. These correspond to the WC0 observations. We note that in SM, a
single gate (12) is permanently activated and no dynamic gate assignment is
performed.

The CCDs also contain circuitry that enables the injection of charge into the
first line of pixels (at the left side of the devices depicted in
Figure~\ref{fig:smAf}), which then transfers through the detector at the normal
rate. Periodic short bursts of injected charge are used to both mitigate and
calibrate the effects of radiation-induced charge transfer inefficiency (CTI) in
the image section of the CCD \citep{2016A&A...595A...6C}.
\subsection{Specification of independent calibration units}
The SM and AF observations made by \textit{Gaia} split naturally into different
subsets over which the PLSF is calibrated independently. Each distinct
combination of field of view, CCD, TDI gate and window class corresponds to one
independent calibration unit.
There are 248 calibration units that correspond to 1D observations for which the
LSF is calibrated; these cover the 62 AF CCDs, two window classes (WC1 and WC2)
and the two fields of view.
In contrast, there are 1020 calibration units that correspond to 2D observations
for which the PSF is calibrated. These cover the 62 AF CCDs, one window class
(WC0), eight TDI gates (4, 7--12, 0) and two fields of view, plus an additional
28 calibration units corresponding to the 14 SM CCDs and two window classes
(WC0 and WC1).
This leads to a total of 1268 calibration units for the PLSF calibration.
\section{Description of the PLSF models}
The task of modelling the PSF for a telescope or image is a classic problem in
astronomy and one for which various standard software packages already exist
(e.g. DAOPHOT \citep{1987PASP...99..191S}, PSFEx \citep{2011ASPC..442..435B}).
However, the extreme requirements for Gaia centroiding accuracy, the unusual
PSF dependences and the highly windowed, undersampled and marginalised data
motivate the development of a dedicated PLSF model and calibration pipeline
tailored to the unique needs of \textit{Gaia}.
\subsection{Basics}
An important point to note is that the true intrinsic or instrumental PSF, which
is the 2D distribution of flux in the focal plane, is never directly observed.
Instead, what is observed and calibrated is the `effective' PSF
\citep{anderson2000} which accounts for the pixelated nature of the image, as
well as a few other sources of smearing such as charge diffusion and differences
between the image motion and charge transfer rate during TDI operation.
As such, the PLSF models used to calibrate the effective PSF must satisfy a
number of mathematical requirements.
They must be continuous in value and in the first derivative.
They must also have a shift invariant sum, that is to say, if either model is
sampled over all space at a set of points one pixel apart, the sum must be
invariant to the sub-pixel location at which the samples are drawn. This
expresses the physical constraint that the total number of photoelectrons
received from a source is independent of the sub-pixel location of the source,
which is a good approximation for back-illuminated CCDs with high fill-factors.
Enforcing this constraint in the PLSF models avoids introducing small
photometric biases as a function of the sub-pixel position.
One complication is that \textit{Gaia} observes only a finite region of the
PLSF due to the use of windowing, and this contains only a fraction of the
total flux (the `enclosed energy fraction', EEF). Over such a region, the
shift invariant sum property does not strictly hold. The PLSF models cannot
account for the EEF, and instead this effect is corrected in the photometric
calibration.

The PLSF models must also be normalised, over one dimension in the case of
the LSF and over two dimensions in the case of the PSF. A subtle issue that
arises here is that due to the finite window extent in the AC direction the LSF
model will underestimate the flux of a star relative to the PSF because the LSF
model fails to account for the flux falling outside of the window in the AC
direction. This AC flux loss is instead accounted for in the photometric
calibration
\citep{EDR3-DPACP-117}.
Another physical constraint is that the true PSF is positive everywhere, however
this is not enforced and as such the calibrated model, fitted to noisy
observations, can be negative in places. Similar to negative parallaxes, this
does not necessarily indicate a problem and simply indicates that the true PSF
value is likely to be small.
\subsection{Formulation of the 1D LSF model \label{sect:lsf}}
The model for the LSF has not changed significantly since DR2 (although the
calibration has--see later), and is described extensively in the associated
documentation \citep[][section 2.3.2]{2018gdr2.reptE...2H} and in several
technical notes, in particular \citet{LL:LL-046, LL:LL-084, LL:LL-088,
LL:LL-089}.

\begin{figure*}
\centering
  \includegraphics[width=0.23\textwidth]{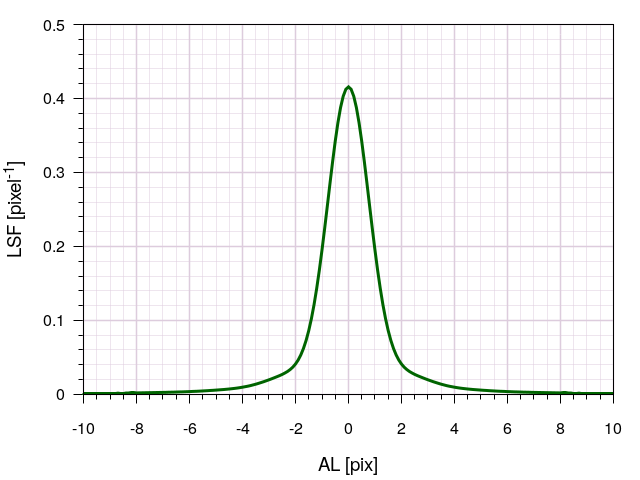}
  \includegraphics[width=0.23\textwidth]{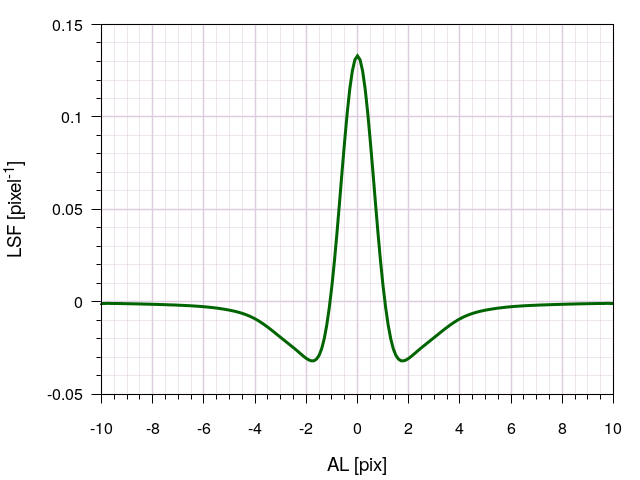}
  \includegraphics[width=0.23\textwidth]{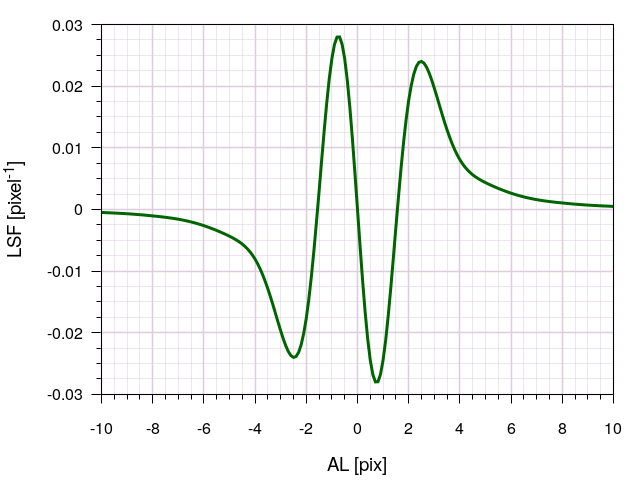}
  \includegraphics[width=0.23\textwidth]{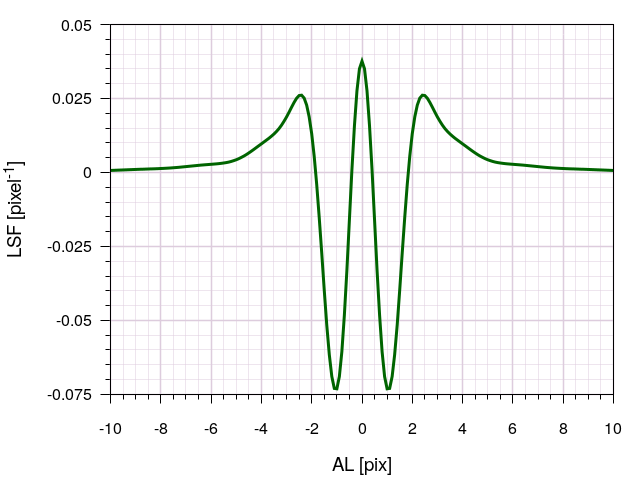}
  \caption{Mean LSF $H_0$ (left) and the first three basis components  
  $H_1$--$H_3$, used to model the LSF for both telescopes. These are obtained
  from simulations, with the odd or even symmetry arising from the inclusion of
  reflected wavefront error maps (see text).}
    \label{fig:lsfbases_0_3}
\end{figure*}

To recap, the LSF $L(u)$ is constructed as the linear combination of a
mean profile $H_0$ and a weighted sum of $N$ basis components $H_n$, where
\begin{equation}\label{eq:lsf}
L(u) = H_0(u-u_0) + \sum_{n=1}^{N} h_n H_n(u-u_0) \, .
\end{equation}
The AL coordinate $u$ has units\footnote{Also
sometimes expressed in \textit{Gaia} documentation in the equivalent unit of TDI
periods. An important point is that due to the use of TDI mode, there is no
correspondence between the AL coordinate and an absolute location on the
detector; rather, the AL coordinate measures only displacements in the AL
direction.} of pixels.
The notation used here is slightly adapted relative to Equation 10 of
\citet{LL:LL-088}, in order to allow greater clarity when presenting both LSF
and PSF models. In particular, the parameter correponding to a shift of origin
is denoted $u_0$ rather than $h_0$. In EDR3 this is not considered a free
parameter and is set to zero; it will be omitted in the following equations
(though see section~\ref{sec:alshift} for further discussion). The only free
parameters of the model that need to be calibrated are the weights $h_n$, which
are themselves multidimensional spline functions of the source colour and other
parameters. The main task of the LSF calibration is thus to solve for the
parameters of these multidimensional splines, which is explained further in
section~\ref{sect:h_fit}.
The functions $H$ in Equation~\ref{eq:lsf} are represented using an `S-spline'
in the LSF core with a smooth transition to a Lorentzian profile in the wings
\citep[see][section 2.3.2.1]{2018gdr2.reptE...2H}, a formulation chosen
specifically to meet the shift invariant sum requirement.
The particular forms of $H$ are derived from simulations of the optical system
carried out before launch, and described in \citet{LL:LL-084}. Briefly, many
random realisations of the \textit{Gaia} wavefront error map were used to
generate a large set of monochromatic PSFs of various wavelengths, assuming a
Fraunhofer diffraction model of \textit{Gaia}'s optical system and applying a
modulation transfer function to model the smearing effect of TDI mode, pixel
binning and charge diffusion.
The monochromatic PSFs were blended to produce a large set of physically
plausible polychromatic PSFs for different stellar spectral energy
distributions, then marginalised to obtain the AL LSFs. This set was doubled in
size by including the reversed LSFs obtained by reflecting in the AL direction.
A principal components analysis (PCA) was applied to these to determine the mean
LSF and the basis components, resulting in a set of orthogonal functions that
can be used to model the LSF and for which the truncated set gives the lowest
possible RMS reconstruction error among all linear models.
We note that the inclusion of the reflected LSFs has the effect of imposing
odd or even symmetry on the mean LSF and the bases obtained by PCA\footnote{This
does not imply that the calibrated LSF formed by a weighted sum of the bases has
any such symmetry---indeed, the calibrated LSF is quite asymmetric, as can be
seen in Figure~\ref{fig:nuEffVariation}.}, as is evident in
Figure~\ref{fig:lsfbases_0_3}.
These were then
post-processed as described in \citet{LL:LL-088} to ensure that the mean is
normalised such that the integral is unity and the basis components are
normalised such that their integrals are zero. This guarantees that the full
model is normalised to unity regardless of the basis component weights $h_n$.
The resulting discretely sampled functions were then fitted with the S-spline
model described above to obtain the final functions used in the data processing.
The mean LSF and the first three basis components are shown in
Figure~\ref{fig:lsfbases_0_3}.
There are some points to note regarding this procedure. First, the generation of
the basis components via random realisations of the wavefront error map means
they are in principle capable of modelling the LSF of a range of instruments
according to the configuration space spanned by the wavefront error maps.
The number of basis components required to reach a particular RMS reconstruction
error could therefore be reduced by tailoring the bases to the in-flight
\textit{Gaia} instrument via a suitable reduction procedure
\citep[e.g.][]{LL:LL-090}, however this was not found to offer a significant
advantage and was not done for EDR3.
Second, efforts were made to determine an appropriate set of basis components
directly from the data, however due to the presence of noise and the
undersampled nature of the observations the resulting basis components
had larger PLSF reconstruction errors than those derived from simulations.
Third, it is important that all significant instrumental effects are
included in the optical model, as any missing component will result in bases
that cannot fully reproduce the real observations. This is expanded on in
section~\ref{sec:newBases}.
Finally, the normalisation procedure applied to the initial basis components has
the side effect of introducing slight non-orthogonality. This effect is minor
and was found to not compromise the numerical stability of the calibration
pipeline.
\subsection{Formulation of the 2D PSF model \label{sect:gen_shape}}
Before describing the PSF model implemented for EDR3, some historical
context is useful. The original PSF model implemented for \textit{Gaia},
referred to in the documentation as the `AL$\times$AC' model, composed the 2D
PSF (denoted $P(u,v)$) as the outer product of two 1D LSF models that were
calibrated to the AL and AC marginal profiles:
\begin{align}\label{eq:alxac}
P(u,v) &= L(u) \times L(v) \nonumber \\
       &= \left(\sum_{n=0}^{N} h_n H_n(u)\right) \times
\left(\sum_{m=0}^{N} g_m H_m(v)\right) \nonumber \\
       &= H_0(u) H_0(v) + \sum_{n+m>0}^{\substack{n=N\\m=N}}
h_{n} g_{m} H_n(u) H_m(v) \, .
\end{align}
where $v$ denotes positions in the AC direction, and the factors $g_m$ are the
basis component amplitudes for the AC LSF. We note that for brevity the sum has
been expanded to include the mean components, so that $h_0=g_0=1$, and the same
1D functions $H$ are used to model both the AL and AC LSFs.
This was assumed to be a reasonable model for \textit{Gaia}, given that the PSF
formed by Fraunhofer diffraction of a rectangular pupil can be factored into the
product of 1D functions in each dimension.
Unfortunately, this fails to take into consideration the wavefront errors, which
introduce significant asymmetric features that cannot be represented within this
model \citep[see][figs 2.1 and 2.2]{2018gdr2.reptE...2H}. While this
AL$\times$AC model was used in the production of DR1 and DR2, it was clear that
a new formulation was required for EDR3.
\subsubsection{The pseudo-shapelets model}
\begin{figure}[!h!]
\centering
  \includegraphics[width=0.45\textwidth]{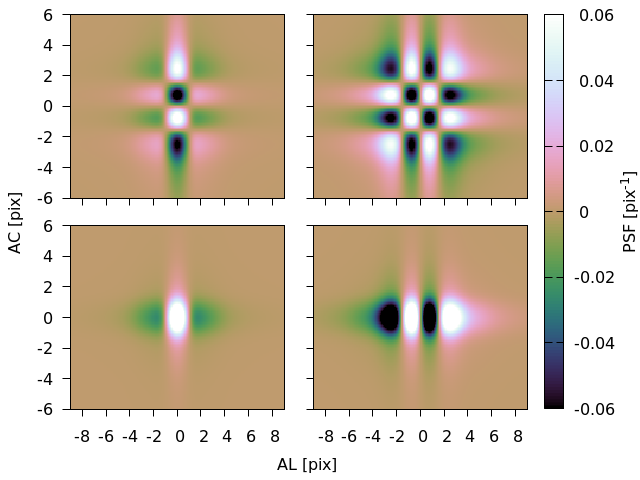}
  \caption{Selection of low order pseudo-shapelets basis components, formed
  from outer products of the 1D functions as follows: $H_1(u)H_0(v)$ (bottom
  left), $H_2(u)H_0(v)$ (bottom right), $H_1(u)H_2(v)$ (top left), and
  $H_2(u)H_2(v)$ (top right). Each pseudo-shapelet has been normalised by a
  different amount in order to better display the structure. Throughout this
  paper we make use of the cubehelix colour scheme introduced in
  \citet{2011BASI...39..289G}.}
    \label{fig:shapelets}
\end{figure}

The PSF model initially developed for EDR3 was based on the shapelets idea
described in \cite{Refregier_2003}, where the PSF is composed as the weighted
sum of 2D basis components (the shapelets) that are generated as the outer
products of orthogonal 1D functions of varying order. In the original paper
these were Hermite polynomials, but for application to the Gaia observations the
1D functions are the same ones used to model the LSF. The resulting PSF model is
referred to as pseudo-shapelets, in light of the fact that the 1D functions are
different.
The full pseudo-shapelets PSF model is then
\begin{equation}\label{eq:shapelets}
P(u,v) = H_0(u)H_0(v) + \sum_{n+m>0}^{\substack{n=N\\m=N}}
h_{nm} H_n(u) H_m(v) \, .
\end{equation}
This is similar to the AL$\times$AC model (Equation~\ref{eq:alxac}), with the
generalisation $h_nh_m \rightarrow h_{nm}$ that allows much greater freedom in
the model to reproduce asymmetric features.
A selection of the low order pseudo-shapelets basis components, formed from the
outer products $H_nH_m$, are presented in Figure~\ref{fig:shapelets}.

While this model achieves good reconstruction of the PSF, it has the major
drawback that a very large number of basis components are required. For 20 1D
basis components $H_n$ plus the mean $H_0$, there are 440 2D bases and an
equivalent number of weights $h_{nm}$, which becomes a major computational
challenge when scaled up to the demands of the \textit{Gaia} data processing,
from the point of view of both calibration of the model and sampling it to fit
observations. In addition, many of the 2D basis components have very
low importance, and there is no way to rigorously rank them to form a truncated
set. Investigation of the principal components present in the real \textit{Gaia}
PSF revealed two important features. First, the observed principal components
have significant asymmetric structure that cannot be well modelled by any
individual pseudo-shapelet. Second, the dimensionality of the real PSF is
significantly lower than 440. These observations motivated the development of
the compound shapelets model that is described in the next section.
\subsubsection{The compound shapelets model}
\begin{figure*}
\centering
  \includegraphics[width=0.23\textwidth]{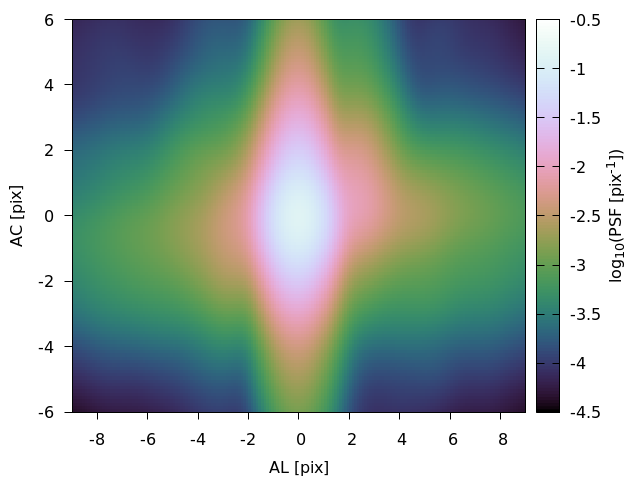}
  \includegraphics[width=0.23\textwidth]{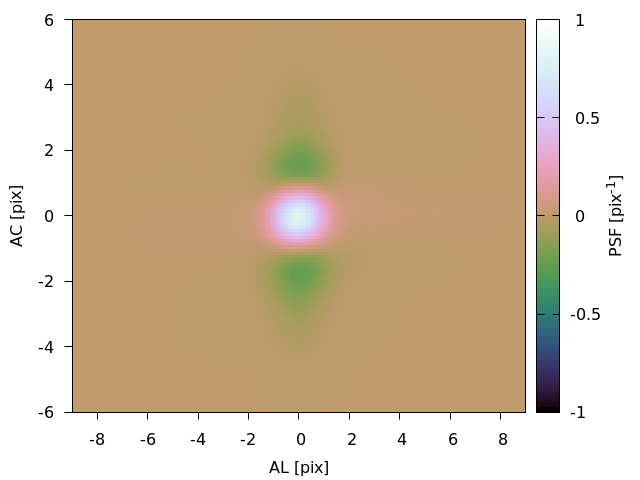}
  \includegraphics[width=0.23\textwidth]{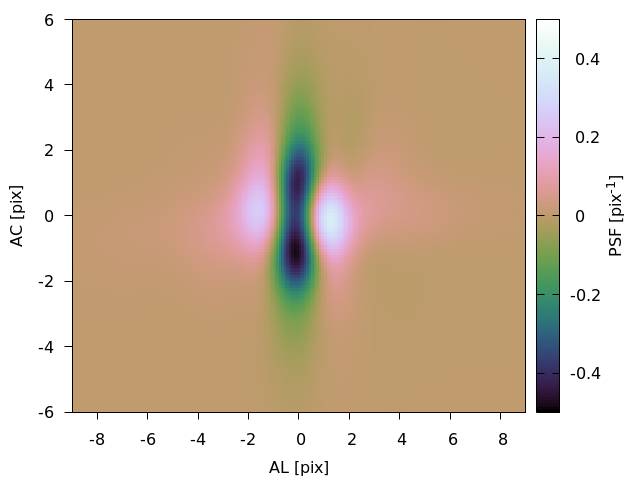}
  \includegraphics[width=0.23\textwidth]{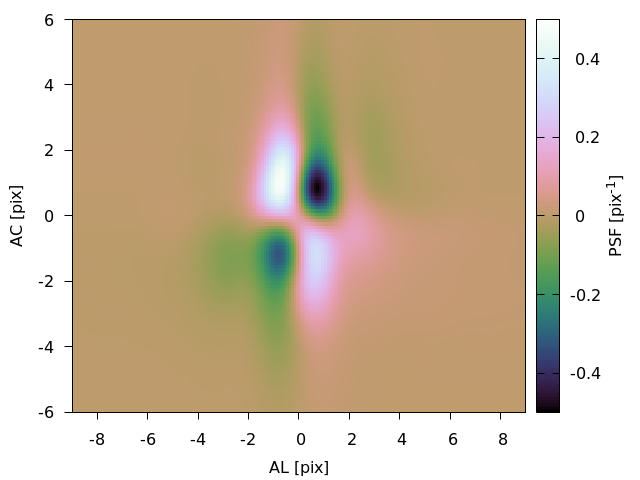} \\
  \includegraphics[width=0.23\textwidth]{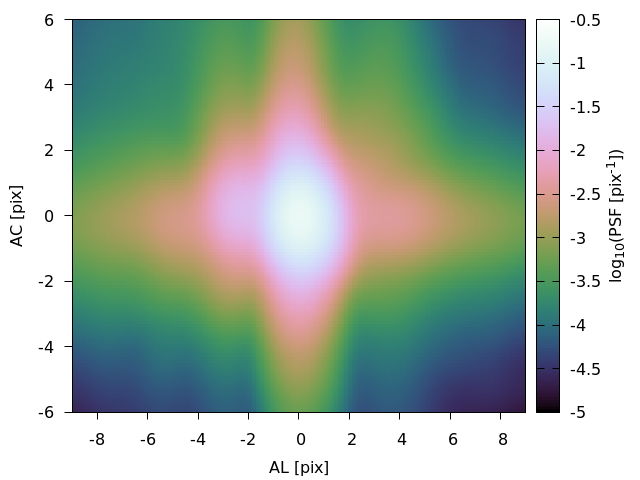}
  \includegraphics[width=0.23\textwidth]{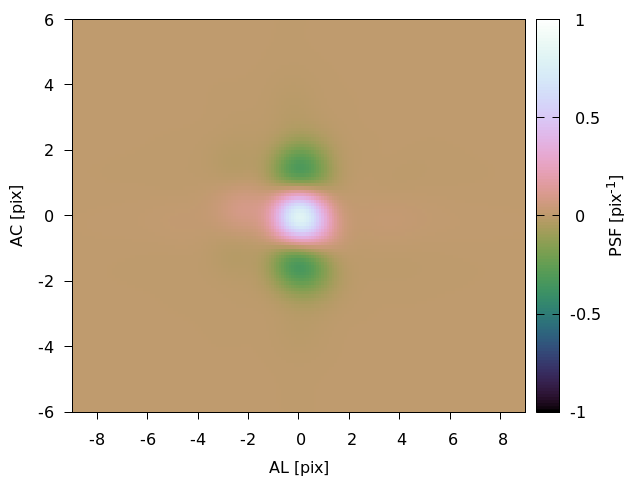}
  \includegraphics[width=0.23\textwidth]{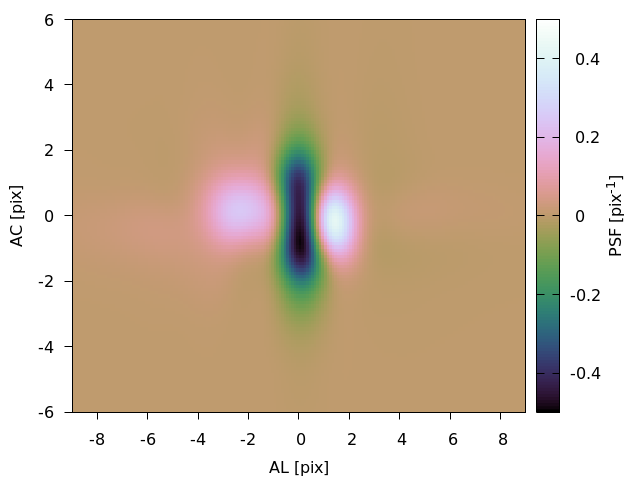}
  \includegraphics[width=0.23\textwidth]{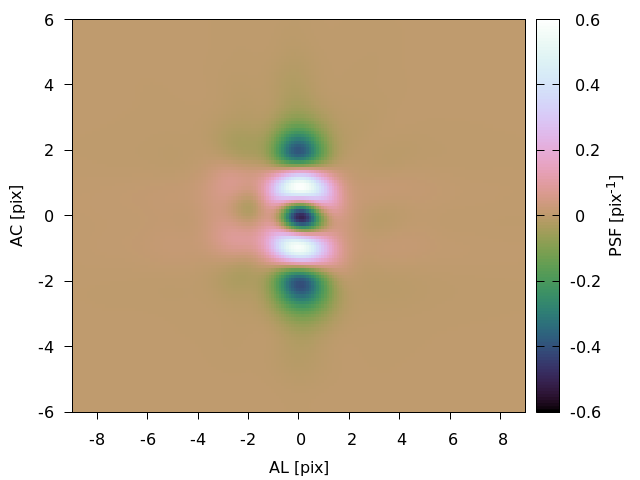}
  \caption{Visualisations of the compound shapelets mean $G_0$ (column 1) and
  first three basis components $G_1$--$G_3$ (columns 2--4) for FOV1 (top row)
  and FOV2 (bottom row). Within each FOV the same components are used to model
  all CCDs. These demonstrate the large differences in the mean PSF between the
  two FOVs, but also the similarity in the low order basis components due to
  the PSFs for the two FOVs being subject to the same major dependences.}
    \label{fig:genshapelets_0_3}
\end{figure*}

The compound shapelets model is based on the pseudo-shapelets model, but rather
than calibrating all the 1D$\times$1D pseudo-shapelets individually, we instead
calibrate fixed linear combinations of them that are constructed to model the
principal components of the observed \textit{Gaia} PSF. Each fixed linear
combination of pseudo-shapelets provides one full 2D basis component that is
referred to as a compound shapelet. The compound shapelets are computed by first
calibrating the pseudo-shapelets model over a large set of \textit{Gaia}
observations that span the whole focal plane and a wide range of mission time,
in order to sample a wide range of instrument states.
The resulting calibrations are then post-processed using the algorithm described
in \citet{LL:LL-090}. This algorithm computes linear combinations of the input
basis components (the pseudo-shapelets) resulting in a transformed set for which
the information is compressed into the leading orders. This provides a minimal
set of 2D bases referred to as the compound shapelets $G$. In terms of these,
the full PSF model is
\begin{equation}\label{eq:genshapelets}
P(u,v) = G_0(u, v) + \sum_{m=1}^{M} g_m G_m(u, v) \,,
\end{equation}
where
\begin{equation*}
G_0(u, v) = H_0(u)H_0(v) + \sum_{k+l>0}^{\substack{k=N\\l=N}} \beta^{0}_{kl}
H_k(u) H_l(v) \,
\end{equation*}
and
\begin{equation*}
G_m(u, v) = \sum_{k+l>0}^{\substack{k=N\\l=N}} \beta^{m}_{kl} H_k(u) H_l(v) \,.
\end{equation*}
The (constant) matrix $\beta$ defines the construction of the compound
shapelets from the pseudo-shapelets. The only free parameters now are the
weights $g_m$ applied to the compound shapelets; for EDR3 30 basis components
were used. Independent sets were generated for FOV1 and FOV2; the mean and first
three bases for both FOVs are depicted in
Figure~\ref{fig:genshapelets_0_3}.

%

\subsection{Modelling of the major PLSF dependences \label{sect:h_fit}}
The observed PLSF exhibits significant variation with source colour, position
in the focal plane and certain other observation parameters that must be
incorporated into the model. There is also large variation in time due to the
evolving instrument state. As described in section~\ref{sec:timedep} we handle
the time dependence in a different manner to the rest of the PLSF
dependences, and it will be omitted in the following description.

All of the major PLSF dependences are ultimately modelled empirically by
appropriate weighting of the PLSF basis components. The weight factors $h_n$
and $g_m$ are represented by multidimensional spline functions of the
observation parameters, each of which is chosen to adequately parameterise the
corresponding dependence.
The particular spline implementation used is that described in
\citet[appendix B]{2007ASSL..350.....V}, extended to multiple dimensions
according to the number of observation parameters included in the model.
The set of observation parameters chosen and the spline configuration for each
dimension define the configuration of the PLSF model, with the parameters of
the model that need to be calibrated being the coefficients of the spline
functions.
For EDR3 we selected the following observation parameters for inclusion in the
PLSF model. These were chosen to represent the largest dependences present.
We note that source flux is not included in the parameterisation of the PLSF
for EDR3; this is discussed in section~\ref{sec:magDep}.
\subsubsection{Source colour}
Because of \textit{Gaia}'s broad $G$ band combined with wavelength-dependent
diffraction within the optical system, the PLSF profiles of stars of
different spectral type show large variation. The source colour is parameterised
by the `effective wavenumber', denoted $\nu_\text{eff}$, which is calculated as
the photon-weighted inverse wavelength. $\nu_\text{eff}$ was identified as a
suitable parameterisation of the source colour as the chromatic shifts in the
PLSF centroid are expected to be linear in $\nu_\text{eff}$
\citep{LL:JDB-028,2006A&A...449..827B}.
The value of $\nu_\text{eff}$ for each source is calculated from the BP and RP
spectra as part of the photometric processing and is expressed in
$\mu\text{m}^{-1}$.
We note that the astrometric `pseudocolour' \citep{dr2_astro} was not used as
it is much less precise than $\nu_\text{eff}$ calculated from the BP and RP
spectra.
\subsubsection{Across-scan rate \label{sec:acRateDep}}
The nominal scanning law of \textit{Gaia} \citep[see][section
5.2]{2016A&A...595A...1G} induces a periodic (\textasciitilde6~h) variation in
the rate at which stellar images drift across the CCDs in the AC direction during
integration (perpendicular to the TDI direction). The modulation is
sinusoidal, and centred roughly on zero with an amplitude of approximately
177~mas~s$^{-1}$ or 1 AC pixel per second.
The AC rate in angular units is denoted $\dot{\zeta}$ (the time
derivative of the AC field angle $\zeta$), however for the purposes of the PSF
model we transform to units of pixels per second by dividing by the nominal AC
angular pixel scale of 176.8~mas~pixel$^{-1}$. In these units the AC rate is
denoted $\dot{\mu}$. Although the in-flight pixel scale differs from the nominal
value (by a different amount for each telescope), the nominal value is
sufficiently close that a calibrated value is not required here.

The systematic difference between the image motion and the motion of the
integrating charge packet causes a broadening of the observed PSF in the AC
direction. For observations with the longest integration time (4.42 seconds,
when no CCD gate is activated), the amplitude of the broadening is around
4.5 pixels.
The strength of the effect varies enormously with the CCD gate length due to the
differing integration times (see section~\ref{sec:instrument}), such that for
gates 10 and shorter (corresponding to integration times $\leq1$ second)
the effect is negligible and the dependence is disabled in the model.

In principle, the effect of the broadening on the PSF should be invariant to the
sign of the AC rate, and in \textit{Gaia} DR2 the absolute value of the AC rate,
$|\dot{\mu}|$, was used in the PSF model. However, it has since been discovered
that small rotational misalignments of the CCDs cause the zeropoint of the
broadening to be offset slightly from $\dot{\mu}=0$. This phenomenon is known as
`native AC rate', and it induces a small asymmetry in the effect on the PSF that
is nevertheless significant. This is accounted for in EDR3 by using the true
value of $\dot{\mu}$, preserving the sign.

One important point to note is that because the broadening is strictly in the AC
direction, the effect manifests only in the PSF observations and as such
the LSF has no dependence on the AC rate.
Although the scan law induces an analogous modulation in the AL image rate, it
has a much smaller amplitude and the models developed for EDR3 assume that
the AL rate matches the parallel charge transfer rate exactly such that no
equivalent broadening effect is present in the AL direction (although see
section~\ref{sec:psfAcRateIssue} for some important consequences of this).
\subsubsection{Across-scan position}
The variation in the PLSF with position in the focal plane is to a large
extent handled by calibrating each device independently.
Within each device, the residual spatial variation manifests only in the AC
dimension---because of the TDI mode in which \textit{Gaia}'s CCDs are operated,
the AL variation is marginalised and not directly observed.
The AC coordinate on the CCD is denoted $\mu$, and is a continuous value
running from 13.5 to 1979.5 across the AC extent of the CCD image area.
We note that in the geometric instrument calibration model
\citep{EDR3-DPACP-128},
the dependence on AC position is not continuous but split into nine
segments that coincide with the CCD stitch blocks \citep[arising from the
manufacturing process; see][figure 5]{2016A&A...595A...1G}. However, the PLSF
is expected to vary smoothly with $\mu$, and we model the dependence with a
continuous function.
\subsubsection{Model configuration}
For EDR3 the spline configurations for all fitted observation parameters are
presented in Tables~\ref{tab:lsfpsf_1d_config}--\ref{tab:lsfpsf_2d_sg_config}.
The configuration for the PSF model is different according to the CCD gate
length, as explained above.
%
\begin{center}
\captionof{table}{Configuration for 1D LSF calibrations}
\label{tab:lsfpsf_1d_config}
\begin{tabular}{llcccc}
    \toprule
    Parameter & Units & Min & Max & Order & Knots \\
    \midrule $\nu_\text{eff}$  & $\mu\text{m}^{-1}$          & 1.24
    & 1.72 & 3     & [-] \\
    $\mu$ & pixels             & 13.5      & 1979.5    & 3     & [-]
    \\ \bottomrule
\end{tabular}
\end{center}
\begin{center}
\captionof{table}{Configuration for 2D PSF long gate (11, 12, 0) calibrations}
\label{tab:lsfpsf_2d_lg_config}
\begin{tabular}{llcccc}
    \toprule
    Parameter & Units & Min & Max & Order & Knots \\
    \midrule $\nu_\text{eff}$  & $\mu\text{m}^{-1}$          & 1.24
    & 1.72 & 3     & [-] \\
    $\mu$ & pixels             & 13.5      & 1979.5    & 3     & [-]
    \\
    $\dot{\mu}$    & pix$\cdot$s$^{-1}$ & -1.0    & 1.0     & 3  
    & [-] \\ \bottomrule
\end{tabular}
\end{center}
\begin{center}
\captionof{table}{Configuration for 2D PSF short gate (4-10) calibrations}
\label{tab:lsfpsf_2d_sg_config}
\begin{tabular}{llcccc}
    \toprule
    Parameter & Units & Min & Max & Order & Knots \\
    \midrule $\nu_\text{eff}$  & $\mu\text{m}^{-1}$          & 1.24
    & 1.72 & 3     & [-] \\
    $\mu$ & pixels             & 13.5      & 1979.5    & 3     & [-]
    \\ \bottomrule
\end{tabular}
\end{center}
Each of the parameters $\nu_\text{eff}$, $\mu$, and $\dot{\mu}$ have typical
values that are orders of magnitude different, and to improve numerical
stability of the model they are normalised internally to the [{-1}:1] range.
We note that the spline configurations for all three families of model are very
simple, employing single-piece (no knots) third-order (quadratic) polynomials in
each dimension.
During development of the model with early versions of the \textit{Gaia}
pipelines and associated auxiliary calibrations, it was found that more complex
configurations did not offer significant improvements to the PLSF
reconstruction and could not be justified.
This will likely be revised for future data releases as the data processing
becomes more refined.

The multidimensional spline implementation adopted for use in this work has a
number of free parameters $N_{\text{par}}$ given by the order $n_d$ and number
of knots $m_d$ in each dimension $d$, according to
\begin{equation*}
N_{\text{par}} = \prod_{d=1}^{D} (n_d + m_d) \, .
\end{equation*}
The PLSF basis components of different order all use the same spline
configuration. Considering the number of basis components used in each model,
the total number of free parameters in the PLSF models are listed in
Table~\ref{tab:params}.
\begin{center}
\captionof{table}{Total number of free parameters for the LSF and PSF models
($N_{\text{total}}$), which depends on the number of parameters per basis
component $N_{\text{par}}$ and the number of basis components
$N_{\text{bases}}$.}
\label{tab:params}
\begin{tabular}{lccc}
    \toprule
    Model & 
    $N_{\text{par}}$ &
    $N_{\text{bases}}$ &
    $N_{\text{total}}$ \\
    \midrule 1D LSF & 9 & 25 & 225 \\
    2D PSF (long gates) &  27 & 30 & 810 \\
    2D PSF (short gates) &  9 & 30 & 270 \\ \bottomrule
\end{tabular}
\end{center}
\subsubsection{Time dependence \label{sec:timedep}}
The time dependence in the PLSF is very significant due to the changes in
contamination level and telescope focus throughout the mission. The evolution
over time is irregular, with large variations that are both smooth, during
quiescent periods, and discontinuous, at decontaminations and refocuses.
For these reasons it is impractical to calibrate the time variation in the same
manner as the other dependences, by adding another dimension to the observation
parameter space. Instead, the PLSF calibration is performed independently
over 0.5-revolution steps, and the evolution of the calibration in time is
solved incrementally by merging the independent calibrations using a square root
information filter \citep{Bierman_1977} implemented using Householder
orthogonal transformations.
A square root information filter is simply a method to solve the recursive
least squares problem in a manner that is particularly numerically stable,
because it does not square the design matrix to form the normal equations.
It is similar to a Kalman filter but without the prediction step.
This technique was pioneered during the Hipparcos
data reduction \citep[see][appendices C and D]{2007ASSL..350.....V}, and is
referred to as a `running solution'.
An exponential decay constant of $\lambda=80^{-1}$ revolutions is used to
down-weight older solutions, which has the combined effects of smoothing out noise, enabling
poorly constrained solutions to converge and allowing slow variations in time
to be tracked. The filter `lag' is eliminated by calibrating forwards and
backwards in time then merging the solutions.
The end result is that the solution at time $t_0$ is a weighted least squares
estimate of the calibration parameters, where the statistical weight of the
contributing data at time $t$ has been reduced by a factor of $\exp({-\lambda}
|t-t_0|)$.
The running solution is capable of tracking gradual changes in the PLSF
profile but tends to smooth over discontinuities. For this reason, the solution
is manually reset (the exponential weight function is truncated) at
discontinuous changes in the PLSF calibration, such as at decontaminations
and refocuses. Over the EDR3 time range there are five such events that require
resets of the calibration for one or both telescopes. A list of these is
presented in Table~\ref{tab:lsfpsf_resets}.
The largest disturbances are the decontaminations, of which three occured during
the EDR3 time range (a further three occurred during the commissioning phase,
and the first in the EDR3 time range is number four).
\begin{center}
\captionof{table}{Resets of the PLSF calibration} \label{tab:lsfpsf_resets}
\begin{tabular}{llcc}
    \toprule
    OBMT [rev] & Event               & FOV1 & FOV2 \\ \midrule
    1329.00    & Decontamination \#4 & \checkmark    & \checkmark \\
    1443.96    & Refocus             &               & \checkmark \\
    2342.00     & Decontamination \#5 & \checkmark    & \checkmark \\
    2574.65    & Refocus             & \checkmark    &  \\
    4124.00     & Decontamination \#6 & \checkmark    & \checkmark \\
    \bottomrule
\end{tabular}
\end{center}
\section{Calibration pipeline}
The PLSF calibration pipeline developed for EDR3 involves a series of
procedures that ultimately solve for the parameters of the PLSF models
over the whole mission time and for all calibration units. The procedures vary in
complexity and differ according to whether or not they can be parallelised in
time, which impacts the implementation and execution plan. Automated validation
algorithms are used to ensure that the \textasciitilde$10^7$ different
solutions meet some predefined quality criteria that guarantee their fidelity.
The products of the pipeline are distributed to downstream consumers within
the \textit{Gaia} Data Processing and Analysis Consortium who require the
PLSF calibration for various higher level data processing tasks. In this
section, we describe the calibration pipeline stages and discuss some important
aspects of the design.
\subsection{Observation preprocessing}
The PLSF pipeline follows the self-calibration principle of the \textit{Gaia}
data processing \citep[][section 3.1]{2016A&A...595A...1G}, which means in
practice that the observations used to calibrate the PLSF are a subset of the
regular science observations and no special calibration data is required.
However not all observations are suitable for use in the PLSF calibration,
and those that are must be carefully selected and prepared. In this section we
describe these preprocessing steps.
\subsubsection{Eligibility of observations}
Each transit observed by \textit{Gaia} provides observations in nine (CCD row 4)
or ten (CCD rows 1--3,5--7) successive CCD strips, and thus may be eligible for
use in calibrating up to ten independent PLSF calibration units. However, not
all observations are suitable, and we define eligibility criteria at both the
transit-level and the strip-level.
Transits that are eligible for use in the PLSF calibration must have a valid
$\nu_\text{eff}$, a valid astrometric solution, and an astrometric excess noise
below 0.5 mas.
These imply that the transit must have been successfully cross-matched to a
known \textit{Gaia} source \citep[as described in][]{EDR3-DPACP-124}. As the
PLSF pipeline is one of the first to run in the data processing cycle (after
the cross-match and several lower level CCD calibrations), the photometry and
astrometry necessarily come from the previous data processing cycle,
specifically from the outputs of the PhotPipe and AGIS systems
\citep[see][]{EDR3-DPACP-117,EDR3-DPACP-128}.
This risks propagating systematic errors from one cycle to the next,
and during the processing for EDR3 an additional iteration with AGIS was
performed in order to mitigate this.

We note that the photometric quantity $\nu_\text{eff}$ is the mean value and not
the epoch value computed from the individual transit.
The mean value is more precise, although it may be inaccurate for sources
that have significant variability. Most of the variable sources will be
removed as outliers during later stages of the PLSF calibration pipeline.
%
The $\nu_\text{eff}$ is needed in order to calibrate the colour dependence in
the PLSF, and provides one of the observation parameters (see
section~\ref{sect:h_fit}).

The astrometric solution is used in conjunction with the attitude and geometric
instrument calibration to predict the location of each source in the observed
pixel stream, and to supply values for the AC rate of the observation. The
predicted locations are used to align the set of observations that are
eventually used to solve the PLSF calibration.
We note that this is a significant departure from more traditional PSF
calibration methods, which would refine an initial empirical estimate of the
source location iteratively with the PSF solution. There are a number of
important reasons for this.
Firstly, the \textit{Gaia} observations are rather undersampled, such that
most empirical location estimators are biased and risk introducing systematic
errors in the PLSF calibration.
Secondly, there is a close coupling between the PLSF calibration and the
global astrometric solution computed within AGIS (which includes the attitude
and geometric instrument calibration). Using predicted locations allow for
consistency between the locations defined within AGIS and those adopted by the
PLSF.
Finally, the use of predicted locations allows effects that shift the PLSF
centroid without any change in the source location, such as chromaticity, to be
calibrated directly in the PLSF. This permits a greater separation of roles
between the PLSF calibration and global astrometric solution.

For transits that pass the transit-level criteria, the following strip-level
criteria are applied, mainly to reject observations that have been compromised
in one way or another by the complex way in which the \textit{Gaia} CCDs are
operated. The sampling and windowing strategy, and its interaction with the
gating, charge injection and other processes that are not synchronised between
the CCD strips, can result in transits for which only a subset of the
strip-level observations are used. Observations in each CCD strip that are
eligible for use in the PLSF calibration must have a single CCD gate (that is
the expected one given the source magnitude), nominal window geometry, no charge
injection present within 50 TDI of the window, a predicted location that is
consistent with the empirical location (derived from the centre-of-flux), and a
profile that is consistent with that of a fixed reference PLSF calibration.
The first of these criteria arises because faint stars that are wholly
or partially coincident in the AL direction with a bright star will be observed
wholly or partially with the gate appropriate for the bright star. As such, they
will have a PLSF profile that is of very low signal to noise, or otherwise
compromised by anomalies associated with the activation and deactivation of the
CCD gate---different samples in the profile will have different integration
times and sample different AL regions of the CCD.
The requirement on the window geometry is necessary due to the truncation of
overlapping windows that occurs when two sources are very close together, such
that the assigned windows are in conflict \citep[see][section
2]{dr1_preprocessing}.
The restriction on the charge injection distance is intended to
exclude observations containing a significant flux contribution from released
charge in the pixels close to the charge injection.
The final two criteria are used to reject observations that have poor predicted
locations or PLSF profiles that are significant outliers. The majority of
these are close pairs that have not been detected as such. The reference PLSF
calibration that is used in this step is discussed further in
section~\ref{sec:refLsfPsf}.
\subsubsection{Selection of observations}
Observations that pass the eligibility criteria then undergo two stages
of selection for use in the PLSF calibration.
In the first stage, all eligible observations are selected from the data stream
uniformly in time up to a limit of 1000 per hour of mission time and per
calibration unit, although only the calibration units corresponding to faint
observations reach this limit. The main purpose of this stage is to throttle the
number of observations and reduce memory overheads. This provides many more
observations than are necessary to formally constrain the PLSF model, however
the need to adequately constrain the solution over the whole observation
parameter space implies that a large number of objects must be made available to
the calibration.

The second stage involves selecting observations uniformly within the PLSF
observation parameter space, particularly in the $\nu_\text{eff}$ and
$\dot{\mu}$ dimensions, in order to fully constrain the model over the whole
range. The distribution of observations within the parameter space is highly
non-uniform and certain regions, such as extreme colours, are very sparsely
populated. It is important that the density of objects used in the calibration
is balanced across the parameter space, to ensure that the PLSF model is well
constrained even for rare types of object.

\begin{figure}
\centering
  \includegraphics[width=0.45\textwidth]{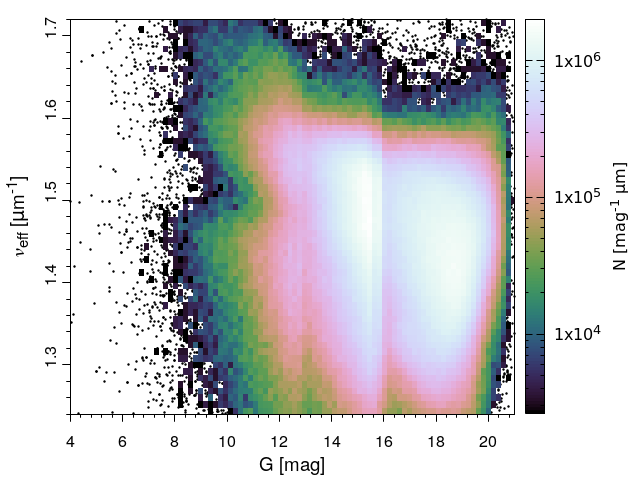}
  \caption{Distribution in the ($G$, $\nu_\text{eff}$) plane of 1604769
  eligible observations that have passed the first stage of selection.}
    \label{fig:nuEffGmag}
\end{figure}

\begin{figure}
\centering
  \includegraphics[width=0.45\textwidth]{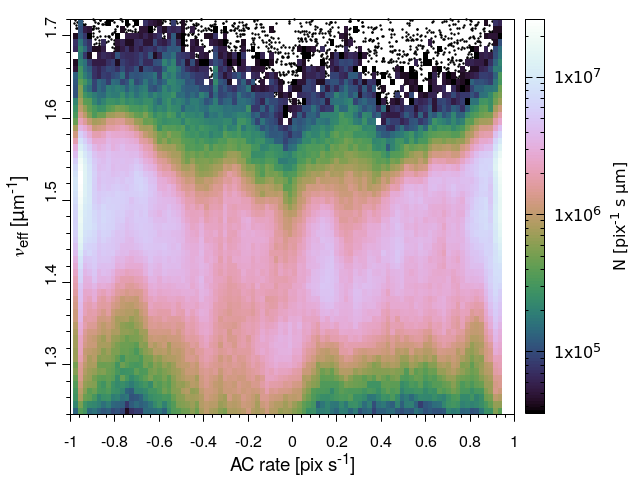}
  \caption{Distribution in the ($\dot{\mu}$, $\nu_\text{eff}$) plane of the
  observations depicted in Figure~\ref{fig:nuEffGmag}. There is a correlation
  between $\dot{\mu}$ and $\nu_\text{eff}$ that evolves over time (see text).}
    \label{fig:nuEffAcRate}
\end{figure}

\begin{figure}
\centering
  \includegraphics[width=0.45\textwidth]{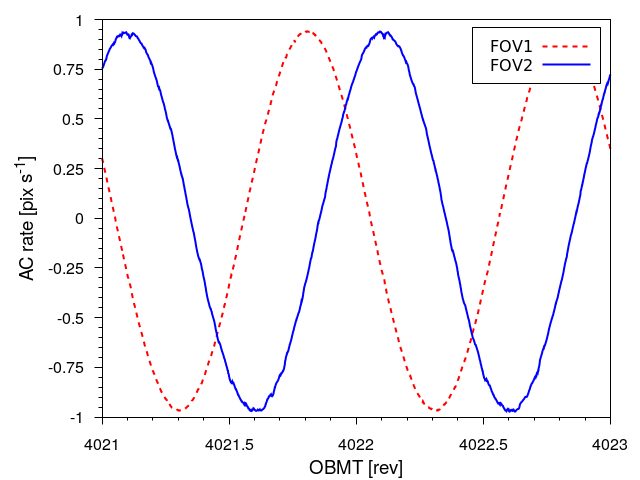}
  \caption{Time variation in AC rate for observations in each telescope.}
    \label{fig:acRateTime}
\end{figure}

To demonstrate this, Figure~\ref{fig:nuEffGmag} shows the distribution in the
($G$, $\nu_\text{eff}$) plane of 1604769 eligible observations that have passed
the first stage of selection. These were observed on CCD row 1 over eight
revolutions from 4020--4028. The step changes at $G=13$ and $G=16$ correspond to
the boundaries between different window classes. The $\nu_\text{eff}$
distribution is highly non-uniform and varies significantly with magnitude.
Figure~\ref{fig:nuEffAcRate} shows how the same observations are distributed in
the ($\dot{\mu}$, $\nu_\text{eff}$) plane. The mean $\nu_\text{eff}$ of sources
on the sky is a function of Galactic latitude, which results in a correlation
between $\nu_\text{eff}$ and AC rate that varies over time. As a consequence of
this, the PSF joint dependence on $\dot{\mu}$ and $\nu_\text{eff}$ cannot be
fully constrained over short time ranges.
Finally, Figure~\ref{fig:acRateTime} depicts the time variation in the AC rate,
which follows the planned scan law closely \citep[see][section
5.2]{2016A&A...595A...1G} with minor departures due to various small
disturbances.
The AC rate varies sinusoidally with a period of 1 revolution, an amplitude of
around 1 pixel per second, and is out of phase between the two telescopes by
$106.5^{\circ}$ (the basic angle). The strong time variation in AC rate implies
that the PSF calibrations with a dependence on this parameter (see
Table~\ref{tab:lsfpsf_2d_lg_config}) require at least half a revolution to be
fully constrained---although the correlation with $\nu_\text{eff}$ and the
sparsity of observations at extreme values of $\nu_\text{eff}$ mean that in
practice a larger time range is required.

Considering these factors, when making the final selection of observations to be
used in the PLSF calibration we define a uniform grid in the observation
parameter space and select one object per bin until either the grid is fully
populated or the observations have run out. This is done separately for each
calibration unit and in steps of 0.5 revolutions, over which the independent
PLSF solutions are computed---see section~\ref{sec:partial}. The grids used
are configured such that the individual 1D LSF solutions can have up to 4000
observations sampled uniformly in ($\nu_\text{eff}$, $\mu$), and the PSF
calibrations can have up to 125 observations sampled uniformly in
($\nu_\text{eff}$, $\mu$, $\dot{\mu}$), although the grids are rarely fully
populated. The difference in total number of observations is due to the fact
that each 1D observation used to constrain the LSF supplies 12 or 18 samples for
use in the fit, whereas each 2D observation used to constrain the PSF supplies
216 (for AF2--9) or 108 (for AF1).
\subsubsection{Preparation of observations \label{sec:obs_prep}}
The observations selected for use in the PLSF calibration must be carefully
prepared. First, observations are converted from analogue-to-digital units (ADU)
to electrons by multiplying by the analogue-to-digital converter (ADC)
gain level.
Then, each observation is corrected to remove the electronic bias, background
and dark signal. Each of these steps relies on auxiliary calibrations of the
associated effect that are computed by the data processing pipelines that run
prior to the PLSF pipeline.
The electronic bias combines the bias prescan level, which is constant for all
samples in an observation although it varies slowly in time \citep[see][section
5.1.1]{dr1_preprocessing}, and bias non-uniformity, which can vary from sample
to sample and has a complex dependence on the CCD readout sequencing as
described in \citet{pemnu}. The background is dominated by stray light due to
the compact, folded design of the optical system, but also includes a minor
component arising from charge release into the pixels behind a charge injection,
although by excluding observations in the first 50 TDI lines following a charge
injection this component is negligible.
The background model has been completely redesigned since DR2 in order to
better reproduce sharp stray light features, and is described in section 3.3.4
of \citet{EDR3-DOC}.

Next, each sample is assessed to determine if it is affected by a range of
defects including saturation, non-linearity, cosmic rays, and various CCD
cosmetic defects such as dead and hot columns. These in turn rely on further
auxiliary calibrations of the CCDs that are determined ahead of the PLSF
pipeline \citep[see][section 2.3.4]{2018gdr2.reptE...2H}. This information is
distilled into a sample mask that defines which of the samples are suitable for
use. Additionally, for the SM observations (which use much longer windows in the
AL direction) we mask off samples in the far wings such that only the central
eight (WC1) or four (WC2) AL samples are used. The motivation for this is that
the samples in the far wings have very low signal to noise and are more likely
to be compromised by secondary sources.

The resulting unmasked samples must then be normalised to match the constraints
of the PLSF models, which require that the flux falling in the unobserved
region outside of the window area (the enclosed energy fraction) is accounted
for---in both the AL and AC direction in the case of the PSF model, and in just
the AL direction in the case of the LSF model.
In principle this could be done by incorporating the photometric calibration in
order to estimate the total instrumental flux of the observation, given the
calibrated magnitude of the source (which is available from the cross-match).
Comparison with the observed number of electrons in the window would then
provide the appropriate normalisation factor. However, due to restrictions
arising from the overall DPAC pipeline design and the formulation of the
photometric calibration this method is not feasible.
Instead, a static reference PLSF calibration is used to fix the enclosed
energy fraction to a nominal value. The reference calibration has no time
dependence, but includes all calibration units and models the colour, AC
position and AC rate dependences so that the enclosed energy fraction has a
physically reasonable value.
This results in observations that are normalised
sufficiently accurately to allow the PLSF solution to stabilise, at the
expense of losing some sensitivity to genuine changes in the enclosed energy
fraction over the course of the mission.
This procedure is somewhat ad hoc, and may be revised substantially for future
data releases as the data processing systems become more refined. This is
discussed further in section~\ref{sec:refLsfPsf}.

The uncertainties on each sample are estimated by combining the shot noise,
readout noise and uncertainties on the associated auxiliary calibrations (bias,
background and dark signal).
An additional contribution is added to account for the uncertainty on the
predicted location of each observation, in order to down-weight observations
with noisier locations. This is done by propagating the location error (in TDI)
to the corresponding error on the (normalised) samples by multiplying it by the
gradient of the PLSF at the location of the sample. Once again the static
reference PLSF calibration is used to estimate this. For 2D observations
both the AL and AC location uncertainties are included. This makes a larger
contribution in steeper regions of the profile where the effects of uncertainty
on the location are more significant.
The various terms that contribute to the estimated sample uncertainties are
unlikely to be fully independent and may in some cases be poorly estimated,
highly non-Gaussian, or correlated between the samples. As a result, the
estimated sample uncertainties are likely to be somewhat biased, which has
implications for the interpretation of the statistics of the PLSF model fit.
However, they provide suitable values for use in weighting the samples used in
the fit.
\subsection{Partial solution \label{sec:partial}}
The solution for the PLSF model parameters is computed independently for each
calibration unit and in steps of 0.5 revolutions. For the 1268 calibration units
and 4152 revolutions covered by EDR3 data, this amounts to 10,529,472 individual
solutions. These are referred to as `partial solutions' because for many
calibration units there are insufficient observations to fully constrain the
parameters.

The PLSF model parameters that must be solved correspond to the parameters of
the spline functions that are used to interpolate the basis component
amplitudes ($h_n$ and $g_m$ in
Equations~\ref{eq:lsf}~and~\ref{eq:genshapelets}).
Using the LSF model as an example, the spline value $h_n$ can be represented as
the inner product of the $P$ spline parameters $\vec{a}^T=[a_1, a_2,
\ldots, a_P]$ and the spline coefficients
$\vec{y}(\vec{o})^T=[y(\vec{o})_1, y(\vec{o})_2, \ldots, y(\vec{o})_P]$ so that
\begin{equation}
h_n = \vec{a}^T \vec{y}(\vec{o}) \, .
\end{equation}
The spline coefficients $\vec{y}$ are functions of the observation
parameters $\vec{o}$, where $\vec{o}=[\nu_\text{eff}, \mu]$ or
$\vec{o}=[\nu_\text{eff}, \mu, \dot{\mu}]$ depending on the PLSF
model (see Tables~\ref{tab:lsfpsf_1d_config}--\ref{tab:lsfpsf_2d_sg_config}).
Each basis component uses the same spline configuration and so has the same
number of parameters that are solved jointly. The full set of parameters can be
represented in a single column vector $\vec{x}$ as follows
\begin{equation}
\vec{x}^T = [\vec{a}_1^T, \vec{a}_2^T, \ldots, \vec{a}_N^T] \, ,
\end{equation}
where $\vec{a}_n$ contains the spline parameters corresponding to basis
component $n$.
The parameter vector $\vec{x}$ can be expressed in a system of linear
equations
\begin{equation}
\label{eq:partial}
A \vec{x} = \vec{b} \, ,
\end{equation}
where the observation vector $\vec{b}$ contains all the samples of the
selected observations, preprocessed as described in section~\ref{sec:obs_prep}
and after further subtraction of the PLSF mean. As the amplitude of the mean
is fixed at 1.0 it is excluded from the fit, and only the amplitudes of the
basis components are solved for by fitting to the sample residuals.
The observation vector $\vec{b}$ can be written as
$\vec{b}^T=[\vec{s}_1^T, \vec{s}_2^T, \ldots, \vec{s}_J^T]$
where $\vec{s}_j$ represents the $K$ samples from observation $j$, and
$\vec{s}_j^T=[s_j^1, s_j^2, \ldots, s_j^K]$ where $s_j^k$ represents the
$k^{\text{th}}$ sample from the $j^{\text{th}}$ observation.
The rows of the design matrix $A$ are composed as
\begin{equation*}
[H_1(u_j^k)\vec{y}^T(\vec{o}_j),
H_2(u_j^k)\vec{y}^T(\vec{o}_j), \ldots,
H_N(u_j^k)\vec{y}^T(\vec{o}_j)]
\end{equation*}
where $H_n(u_j^k)$ represents the value of basis component $n$ at the location
$u$ of the sample $s_j^k$.
Finally, the error on each sample $\sigma_s$ is used to weight the entries in
the observation vector and design matrix, by dividing the corresponding element
in $\vec{b}$ and row in $A$ by $\sigma_s$.
Although the LSF model has been used as an example here, the equations extend
naturally to the PSF model and the same methods are used to solve for the
PSF parameters.

The partial solution is obtained by applying Householder orthogonal
transformations to Equation~\ref{eq:partial} that reduces matrix $A$ to a
particular upper triangular form. For further details, see \citet[appendix
C]{2007ASSL..350.....V} and \citet{Bierman_1977}.
\subsection{Running solution \label{sec:runningSol}}
The partial solutions from independent half-revolution time steps contain noise,
may in some cases not have a unique solution due to fewer observations than
parameters, and in most cases are not well constrained over the whole
observation parameter space due to a lack of objects in certain regions.
In order to reduce noise and improve the constraint, while tracking gradual time
evolution in the PLSF calibration, the partial solutions are combined using
the running solution methodology described in section~\ref{sec:timedep}. In
effect this produces an updated solution for each half-revolution time step,
that is a merger of many earlier and later partial solutions weighted according
to the time difference.
\subsection{Data gaps}
At certain periods throughout the EDR3 time range there may be no observations
available to constrain the partial solution for some or all of the calibration
units. This may happen by chance for bright calibration units for which stars in
the magnitude range are rare. This also happens at times due to anomalies on the
satellite, problems during the downlink of data, or transient issues affecting
the various auxiliary calibrations that are required to prepare the PLSF
inputs. Observations can also be excluded by design, for example during each
decontamination when the collection of science data is halted, and for a short
time afterwards while the instrument is still thermally unstable.

During these periods, the PLSF pipeline still runs but the partial solution
produced is that of the identity solution. This carries no weight in the running
solution, such that the running solution is simply propagated over the gap in
the data with no change. This ensures that there is always a well-constrained
PLSF calibration available, for example to process science observations that
are present but that did not meet the requirements for use in the PLSF
pipeline.
\subsection{Autoqualification and outputs \label{sec:autoqual}}
The scale of the data processing and calibration task, with more than ten
million calibrations each of which solves the amplitudes of tens of basis
components over a two or three dimensional parameter space, necessitates the use
of automated qualification and validation algorithms to monitor the integrity of
the calibration pipeline.
These are under constant review and development as the PLSF models evolve and
our understanding of the data improves.

At the lowest level, these amount to verifying that certain parameters are
within allowed ranges and that undefined values have not entered the
calibration. Gaps in the input observations are detected and checked
against known mission events \citep[e.g.][]{2020MNRAS.497.1826B}. The statistics
of the PLSF fits are tracked to reveal any time ranges or subsets of the
calibration where the models cannot reproduce the observations as accurately as
expected. For every individual calibration, the PLSF solution is inspected at
a range of points within the observation parameter space in order to check its
integrity. Invalid PLSF solutions are those for which the average basis
component amplitudes exceed a threshold, or those for which the reconstructed
PSF or LSF goes significantly negative or has a large number of local maxima.
Some margin is necessary to account for the fact that the PLSF solution is
produced by a calibration to noisy data, and as such may contain minor
unphysical features that do not necessarily indicate a problem with the
pipeline.

Ideally, with PLSF models incorporating all known physical effects, accurate
auxiliary calibrations and properly characterised error distributions, all
PLSF solutions would pass autoqualification and be approved for use. However,
with EDR3 this situation has not yet been reached, and there are some subsets of
the calibration units that have not been adequately calibrated, as explained in
section~\ref{sec:sol_copying}.
Any solution that fails autoqualification must be replaced with a suitable
alternative, which invariably means either copying a solution from the same
calibration unit at a different time, or copying the solution from a different
calibration unit for which the PLSF solution is expected to be similar.

For every half-revolution period, the full set of calibrations for the whole
focal plane are compiled into a single product for distribution to downstream
consumers within the \textit{Gaia} DPAC. Each such PLSF `library' is roughly
4.9MB in size, rising to \textasciitilde570MB when the covariance information on
the PLSF parameters is included; the total size of the pipeline products is
therefore \textasciitilde4.7TB.
\subsection{Implementation and execution}
The PLSF models and calibration pipeline are implemented in the Java
programming language as part of the \textit{Gaia} DPAC codebase.
Within the overall DPAC architecture the PLSF calibration pipeline forms part
of the CALIPD processing system that is referred to in other DPAC publications.
CALIPD combines the instrument calibrations (CAL), which include the PLSF,
and the IPD. The PLSF calibration
is broken into a series of six modules that are executed in series and which
each perform a different stage of the pipeline: observation preprocessing,
partial solution computation, forwards- and backwards-in-time running solutions,
solution merger, and output assembly and packaging.
The running solution and output assembly tasks must be executed sequentially by
time, whereas the other tasks can be parallelised by time. Tasks can also be
parallelised by calibration unit for further optimisation, with the overall
architecture being tailored towards the execution environment.

\begin{figure}
\centering
  \includegraphics[width=0.45\textwidth]{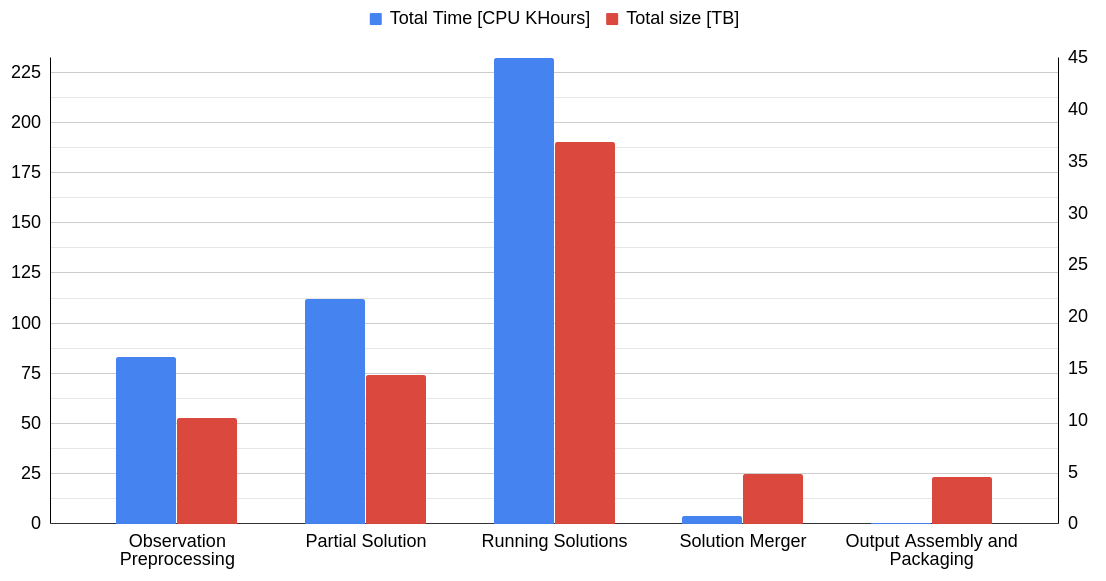}
  \caption{Performance of the PLSF calibration pipeline implemented for  
  EDR3 in terms of the resource consumption (CPU hours and disk space) by
  different stages of the pipeline.}
    \label{fig:lsfPsfExecution}
\end{figure}

The execution of the PLSF calibration is done at the Data Processing
Centre, Barcelona (DPCB), in particular at the MareNostrum supercomputer
hosted by the Barcelona Supercomputing Center (BSC).
Because of the machine design the use of databases is discouraged, and instead a
hierarchical file system was used for hosting the raw data and intermediate
products.
The MareNostrum resources are not exclusively dedicated to \textit{Gaia} and the
access is through job submission to a shared queue. Consequently, the PLSF
calibration has been developed with batch processing in mind.
This decision was largely based on the nature of the majority of DPAC tasks
executed at DPCB, where a lot of data has to be processed by loosely coupled
tasks.
Batch processing is very efficient in processing high volume data, where data is
collected, entered to the system, processed and then results are produced in
batches. The use of partitioning allows multiple jobs to run concurrently,
thus reducing the elapsed time required to process the full data volume.
Special care must be taken in the partitioning of jobs to exploit the
available resources efficiently.
In total, the execution of the PLSF calibration pipeline for EDR3 consumed
around 430,000 CPU hours and required around 66TB of storage for the outputs
and intermediate products. Figure~\ref{fig:lsfPsfExecution} depicts the
distribution of resources among the pipeline stages.
\subsection{Image parameter determination \label{sec:ipd}}
As an aside, it is useful to the reader to briefly describe how the PLSF
calibration is actually used in the \textit{Gaia} data processing. Further
details can be found in section 3.3.6 of \citet{EDR3-DOC}.
Within the overall \textit{Gaia} processing chain, the PLSF models are used
as part of the pre-processing of the raw data that aims to determine, for every
window, the basic observables (or `image parameters') of location (in the pixel
data) and flux. Every window is assumed to contain exactly one point source. 
The image parameters form the primary inputs to both the astrometric and
photometric ($G$ band) calibrations described in other papers. This IPD relies
on numerous auxiliary calibrations in addition to the PLSF.
After debiasing, the window samples are fitted with a model consisting of the
sum of a LSF (or PSF for 2D windows) and a constant background offset. The
PLSF solution is selected according to the FOV, device, CCD gate and time of
the window, and is configured with the ($\nu_\text{eff}, \mu, \dot{\mu}$)
parameters of the observation (a default value is adopted for $\nu_\text{eff}$
if not known).
The image parameters that are solved for are the location (AL only for 1D
observations, or AL and AC for 2D observations), instrumental flux and the
background offset. The background offset is a nuisance parameter that is fitted
in order to better handle sharp features in the stray light variation.
The fitting is performed using a maximum likelihood algorithm described in
\citet{LL:LL-078}.
The adopted `centre' of a star is defined by the origin of the PLSF model,
which itself is aligned with the predicted locations of the observations used to
calibrate the model. These are supplied by the astrometric solution. This
circularity leads to a degeneracy between the PLSF origin and the geometric
part of the instrument calibration (performed within AGIS), which future
data releases will aim to resolve (see section~\ref{sec:alshift}).
The assumption of a single point source is obviously a simplification, and
within DPAC there are subsystems dedicated to the processing of extended objects
and non-single-stars, although these results are not part of EDR3.
\section{Results}
In this section we present some results of the calibration. During the
processing for EDR3, two iterations of the PLSF and AGIS calibrations were
performed in order to improve the convergence and reduce some systematic errors
present in the first iteration. The outputs of the second iteration provided the
inputs used to compute the final EDR3 data products, and it is the results of
this second PLSF calibration that are presented here.

Given the size, complexity and dimensionality of the PLSF calibration
products, it is not feasible to present everything and care is required in order
to distill the results into a meaningful set of analyses.
We present a selection of results from specific subsets of the calibration that
are carefully chosen to demonstrate certain key aspects, and which are
representative of the calibration as a whole.
In many cases the along-scan full-width-half-maximum (AL FWHM) is used to
quantify the image sharpness, reveal gradual evolution in the instrument, and
provide a proxy for the relative astrometric constraint.
\subsection{Time evolution \label{sec:resultsTimeEvolution}}
\begin{figure}
\centering
  \includegraphics[width=0.45\textwidth]{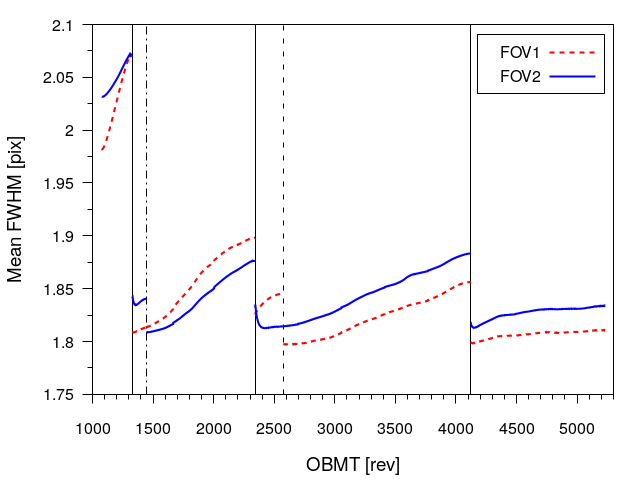}
  \caption{Time variation in mean AL FWHM of the AF2-9 WC1 calibrations, for
  FOV1 (dashed red line) and FOV2 (solid blue line). The vertical lines mark
  decontaminations (solid line) and refocusses of FOV1 (dashed line) and FOV2
  (dot-dashed line). These coincide with resets of the calibration running
  solution.}
    \label{fig:fwhm_dr3}
\end{figure}

In Figure~\ref{fig:fwhm_dr3} we present the mean AL FWHM over all AF2-9 1D
calibrations, for each FOV separately, and over the whole EDR3 time range.
AF1 devices, which use a different window length, are excluded to maintain
consistency in the inputs. The vertical lines mark the times at which the
calibration running solution (section~\ref{sec:timedep}) is reset due to a major
mission event (see Table~\ref{tab:lsfpsf_resets}). These include
decontaminations (solid lines), and refocusses of FOV1 (dashed line) and FOV2
(dot-dashed line).
The time evolution thus corresponds to a gradual degradation in the image
sharpness due to slowly changing instrument focus, punctuated by step changes.
The rate of degradation is higher earlier in the mission: The payload has become
more stable throughout the years, which manifests in other effects such as
slower rates of mirror contamination. After the sixth decontamination at
revolution 4124 the degradation in image sharpness is very modest, and at the
time of writing (revolution \textasciitilde9800) there have been no more
decontaminations and the image quality has been maintained by refocussing alone.

\begin{figure}
\centering
  \includegraphics[width=0.45\textwidth]{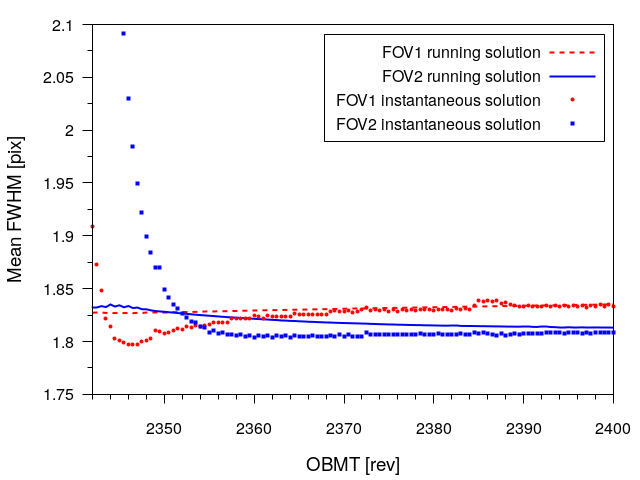}
  \caption{Zoom-in of Figure~\ref{fig:fwhm_dr3} showing the period shortly
  after the fifth decontamination when the instrument is cooling and the
  PLSF is evolving rapidly. In this situation the running solution (solid
  lines) diverges from the partial solution (points).}
    \label{fig:fwhm_decont5}
\end{figure}

After each decontamination \textit{Gaia} takes some time (typically a few tens
of revolutions) to reach thermal stability, and the evolution of the PLSF is
quite rapid during this period. Figure~\ref{fig:fwhm_decont5} shows a zoom-in of
the time range immediately after the reset of the PLSF running solution
following decontamination five at revolution 2342. In this figure the AL FWHM
obtained from the running solution (lines) is compared with that obtained from
the partial solution (points), which provides a calibration of the instantaneous
instrument state.
The rapidly improving image sharpness as the instrument cools cannot be
accurately tracked by the running solution, due to the way in which it merges
solutions over a wide time range. This leads to a systematic difference between
the true PLSF and the calibrated solution for a short period after each
decontamination. Ways to mitigate this in future data releases are discussed in
section~\ref{sec:discRunningSolutionAfterDecontamination}.
\subsection{Colour, AC position, and AC rate dependence}
\begin{figure*}
\centering
  \includegraphics[width=0.95\textwidth]{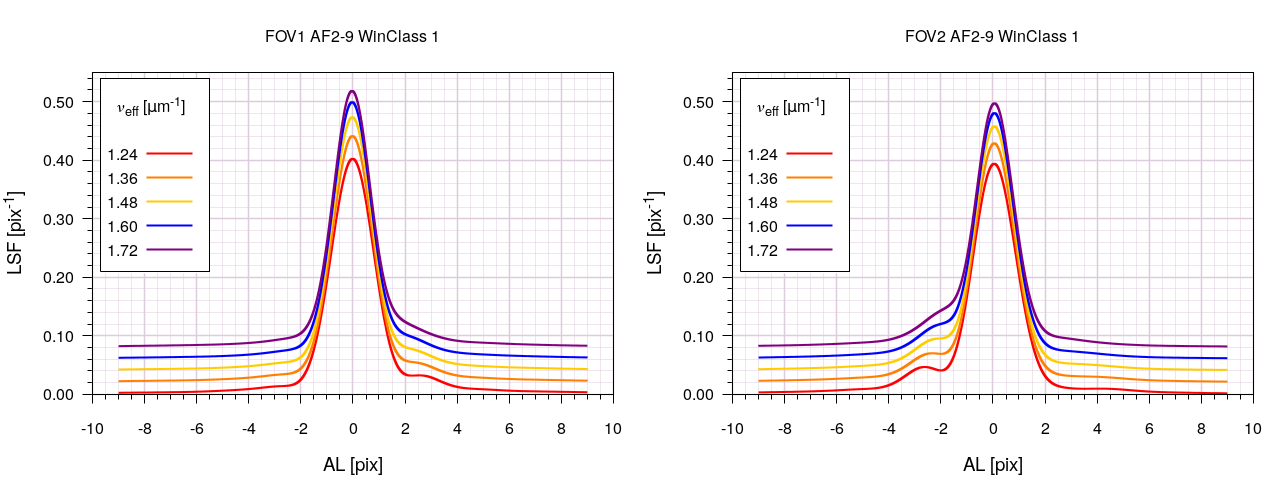}
  \caption{Colour variation for the AF2-9 WC1 LSF for FOV1 (left)
  and FOV2 (right). The profiles have been offset vertically by 0.02 for
  clarity. There is a clear trend with profiles for larger effective wavenumbers
  being sharper with less obvious diffraction features in the wings.}
    \label{fig:nuEffVariation}
\end{figure*}

At each instant in time the PLSF dependences on colour, AC position, and AC
rate are calibrated. The effects of variation in source colour are depicted in
Figure~\ref{fig:nuEffVariation}, where the mean of the calibrated AL profiles
for different values of the effective wavenumber $\nu_\text{eff}$ are shown. The
two FOVs present quite different AL profiles due to the different wavefront
errors between the two telescopes, with the first diffraction peak appearing on
opposite sides of the central maximum. This overall form is present throughout
the entire time range and for the PSF as well. The variation with source colour
manifests through the strong dependence on $\nu_\text{eff}$; as expected,
smaller values of $\nu_\text{eff}$ (which correspond to longer effective
wavelengths) resulting in broader profiles with stronger diffraction features.

\begin{figure}
\centering
  \includegraphics[width=0.45\textwidth]{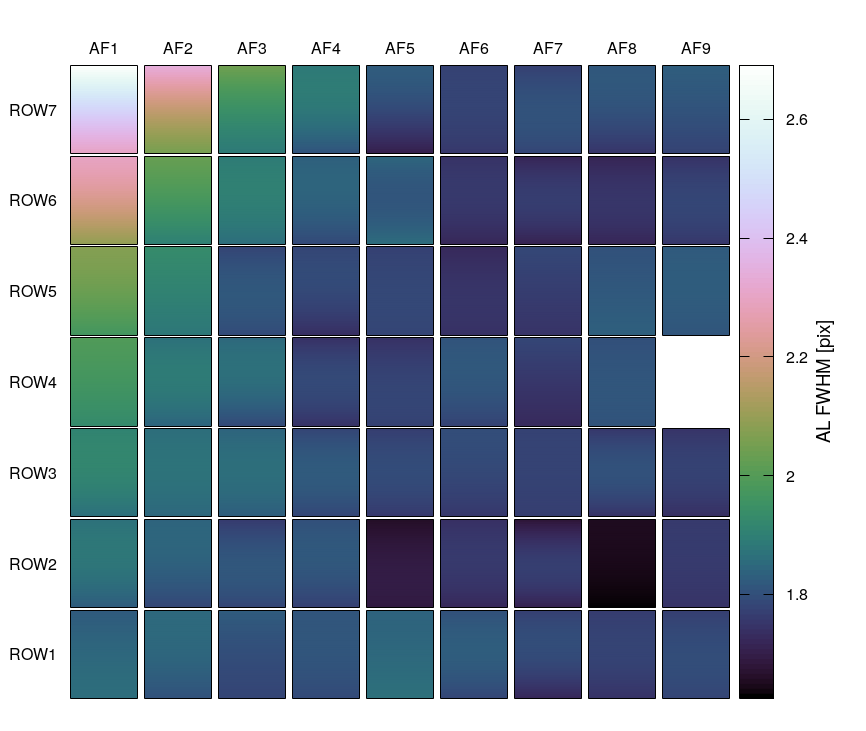}
  \includegraphics[width=0.45\textwidth]{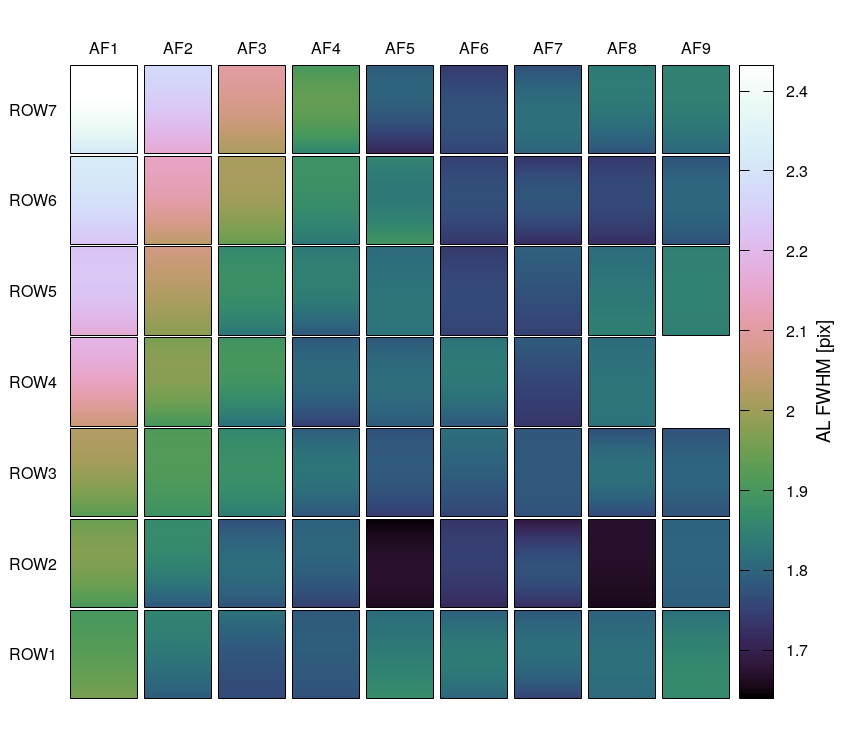}
  \caption{Variation in the AL FWHM with AC position within each CCD and across
  the AF focal plane, for FOV1 (upper panel) and FOV2 (lower panel).
  Obtained from the calibration at revolution 3343 and for
  $\nu_\text{eff}=1.43$$\mu$m$^{-1}$.}
    \label{fig:acPosVariation}
\end{figure}

The dependence on AC position is depicted in Figure~\ref{fig:acPosVariation},
which shows the variation in AL FWHM across the entire AF instrument for each
FOV separately, obtained from the WC1 (LSF) calibrations at revolution 3343 and
for $\nu_\text{eff}=1.43$$\mu$m$^{-1}$.
The 62 CCDs that comprise the AF instrument are shown at true relative size and
position. In principle, the profiles are expected to be sharper closer to the
optical centre and to vary smoothly across the focal plane, and this is indeed
reflected in the calibration, which solves each device independently and does
not enforce these properties \textit{a priori}. For both FOVs there is a clear
degradation of the image sharpness towards the corner of the focal plane
at CCD row 7 and strip AF1,
a feature well established since the commissioning phase
\citep{2014SPIE.9150E..0KB}.
The evolution is not completely smooth
across the focal plane, and some devices show small discontinuities with their
neighbours. These could be due to a general lack of constraint towards the edge
of the CCDs, issues with the auxiliary calibrations, or minor unmodelled
electronic effects that depend on the AC position.
%
%
There are also two CCDs, AF5 and AF8 in row 2, that have 
exceptionally good image sharpness in both FOVs.
Examination of observations in these devices indicates that
the change in PSF shape is genuine and not an artefact of the calibration. The
fact that both FOVs are affected suggests that the root cause lies with
the CCDs and not the optical part of the PSF.
These two devices have also been found to have lower quantum efficiency than the 
other AF CCDs, with a depressed sensitivity at redder wavelengths, which could
explain the difference in PSF.

\begin{figure*}
\centering
  \includegraphics[width=0.95\textwidth]{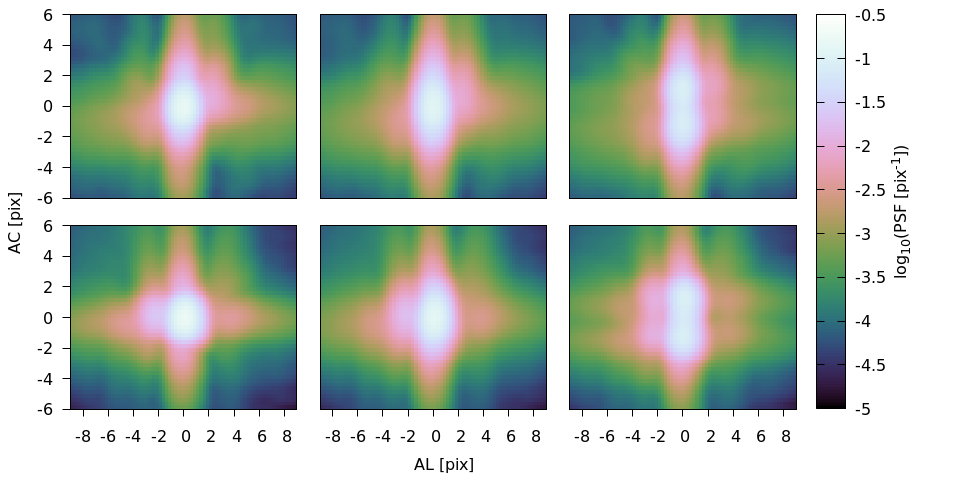}
  \caption{AC rate ($\dot{\mu}$) variation for the FOV1 (top row) and FOV2
  (bottom row) calibrations for CCD row 4, strip AF5, gate 0. The plots
  correspond to AC rate values of 0.0 (left), 0.6 (middle), and 1.0 (right)
  pixels per second. See the text for details.}
    \label{fig:acRateVariation}
\end{figure*}

The variation with AC rate manifests only in the PSF, and leads to a broadening
in the AC direction as the stellar image moves during integration. The strength
of the dependence varies according to the CCD gate, as explained in
section~\ref{sec:acRateDep}, with gate 0 having the strongest dependence. In
Figure~\ref{fig:acRateVariation} we present a selection of calibrated PSF models
for different values of the AC rate (0.0, 0.6 and 1.0 pixels per second, from
left to right) and for FOV1 (top row) and FOV2 (bottom row). These are taken
from the CCD row 4, strip AF5, gate 0 calibration at revolution 4400. They
clearly demonstrate the broadening effect on the PSF for non-zero values of AC
rate. The slight bimodality in the PSFs at high AC rate (right panels) is an
artefact of the calibration that has been studied extensively since the EDR3
calibrations were computed. This is discussed in detail in
section~\ref{sec:psfAcRateIssue}.
\subsection{Invalid solutions \label{sec:sol_copying}}
The autoqualification and validation criteria described in
section~\ref{sec:autoqual} are not met by all subsets of the calibration.
Specifically, the calibrations for SM CCDs and AF WC2 observations
fail various thresholds and have not been qualified for use. It is for this
reason that the SM and AF WC2 calibrations are not included in the results
presented in this section. In both cases, the solutions for these calibration
units are discarded and replaced with qualified solutions from other
calibrations, as explained below.

The SM calibrations for both WC0 and WC1 are complicated by numerous differences
in the data compared with similar AF observations. These include
the use of a fixed CCD gate (12) that does not adapt to the magnitude of the
source (leading to WC0 observations heavily affected by saturation), on-chip
binning (by 2x2 pixels for WC0 and a further 2x2 binning in software to 4x4
pixels for WC1) that results in a highly undersampled PSF, higher ADC readnoise
(of \textasciitilde10.8 e$^-$ compared to \textasciitilde4.3 e$^-$ for AF2-9
strips, see \citet{pemnu}), and the fact that the SM PSF has quite a different
form to AF due to being further from the optimum wavefront location.
The final issue causes problems with the PSF model because the basis components
are tailored towards AF observations. In addition, the PSF model has problems
reproducing gate 12 observations (see section~\ref{sec:psfAcRateIssue}) that
further destabilise the fit.
Many of these issues also have an impact on the various auxiliary calibrations
that the PLSF calibration relies on. In particular, the geometric instrument
calibration for SM is noisier, which results in noisier predicted locations for
each observation.
Each SM PSF solution is therefore discarded and replaced with the solution
from the AF2 strip with the same CCD gate, row, FOV, and mission time. This
choice was judged to be a good compromise between minimising spatial variations
in the PSF while maximising the solution stability.
This procedure is not ideal and major effort has been spent on resolving these
various problems in order to achieve an independent calibration of the SM PSF
for future \textit{Gaia} data releases (see section~\ref{sec:c04}).
However, we also note that in \textit{Gaia} EDR3 the SM observations have not
been used in the astrometric or photometric solutions for sources, and as such
the SM calibrations are of lower importance to AF.

The AF WC2 calibrations suffer mainly from low signal to noise in the
observations. Also, for these faint observations the stray light background is
more significant, and uncertainties in the background calibration affect the
solution to a greater extent.
The solutions for the AF WC2 calibrations are therefore discarded and replaced
with the solution for the WC1 calibration in the same device, FOV and mission
time.
Being coincident in the focal plane, the linear part of the LSF is expected to
be identical between WC1 and WC2, with any differences limited to signal-level
dependent effects that are in any case not included in the LSF model at this
point.
%
%
\subsection{Fit statistics and calibration residuals \label{sec:resid}}
\begin{figure}
\centering
  \includegraphics[width=0.45\textwidth]{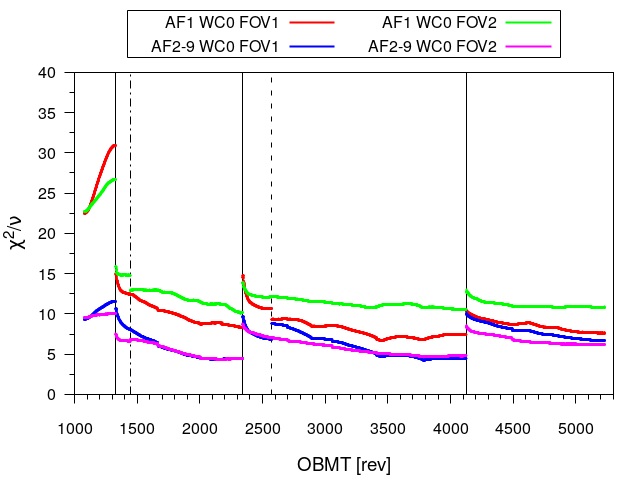}
  \includegraphics[width=0.45\textwidth]{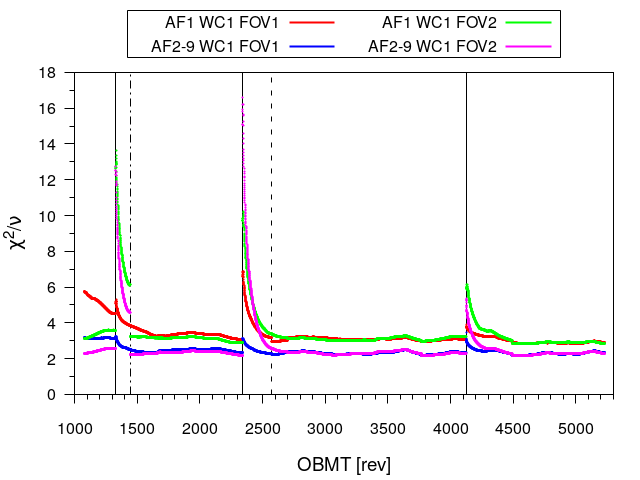}
  \caption{Evolution of the median PLSF model goodness-of-fit statistic
  $\chi^2/\nu$ over the EDR3 time range, as a function of AF CCD strip and
  FOV. The top panel corresponds to WC0 (PSF) calibrations and the bottom panel
  refers to WC1 (LSF) calibrations.}
    \label{fig:gofEvolution}
\end{figure}

%
%
In Figure~\ref{fig:gofEvolution} we present the evolution of the PLSF model
reduced chi-square statistic $\chi^2/\nu$ over the EDR3 time range, as a
function of AF CCD strip and FOV, obtained from the running solution.
For each trend, the median value for the various contributing calibration units
is plotted. This
demonstrates the general stability of the calibration and the similar levels of
goodness-of-fit achieved for the two FOVs, which for the PSF solutions use
different sets of basis components (see Figure~\ref{fig:genshapelets_0_3}).
The goodness-of-fit for the AF1 devices is slightly worse than AF2-9. There are
a number of factors that contribute to this. First, the PSF observations in AF1
are binned on-chip in 1x2 pixels ALxAC, such that the samples used in the fit
tend to have higher signal to noise than AF2-9 and departures from the model are
more significant. Both the LSF and PSF calibrations will be affected to some
extent by the higher electronic read noise in AF1 (8.5e$^-$ versus 4.3e$^-$; see
\citet{pemnu} table 1). Also, the offset non-uniformity part of the electronic
bias cannot for technical reasons be completely removed in AF1
\cite[see][section 3.3]{pemnu}, which leaves uncorrected instrumental signatures
in the observations used in the PLSF fit.
The $\chi^2/\nu$ is also seen to be higher shortly after each decontamination;
this is due to the issue described in section~\ref{sec:resultsTimeEvolution}
where the rapid evolution of the PLSF as the instrument cools cannot be
tracked accurately by the running solution.

\begin{figure}
\centering
  \includegraphics[width=0.45\textwidth]{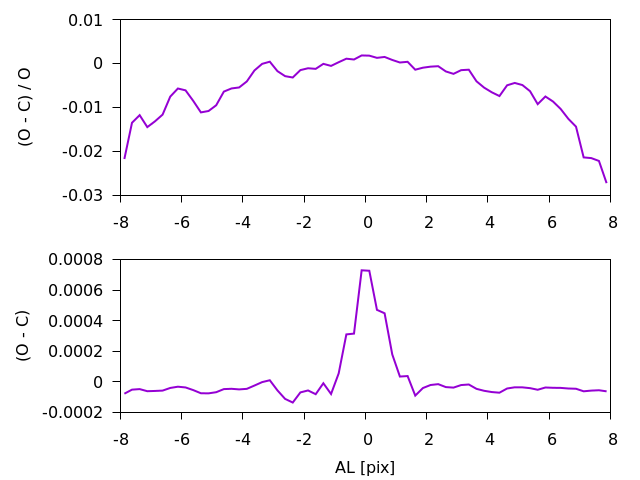}
  \caption{
  Median relative (top) and absolute (bottom) residuals about the model
  for a selection of 1D observations used in the LSF calibration. Observations
  are selected from one calibration unit over a ten revolution period, and span
  the magnitude range $13 \lesssim G \lesssim 16$.}
    \label{fig:lsfResid}
\end{figure}

Figure~\ref{fig:lsfResid} shows an example of the median relative (top panel)
and absolute (bottom panel) residuals about the model for a selection of
observations used in the LSF calibration.
The observations and corresponding model are from a single 1D
calibration unit (FOV2, CCD row 1, strip AF5, gate 0, WC1) over 4020--4030
revolutions, although the results are representative of the whole focal plane.
The observations span the corresponding magnitude range of $13 \lesssim G
\lesssim 16$.
These figures quantify the level of systematic errors present in the LSF
calibration and the degree to which the model can reproduce the observations.
There is clear structure present in the LSF core in the absolute residuals,
suggesting the presence of unmodelled effects that are however rather modest in
terms of their relative size. More significant in relative terms are the
departures in the wings, where the model is systematically larger than the
observations by up to a few percent.
We initially thought that this might indicate systematic errors in the
background estimation, but experiments suggest this is not the case. Instead,
this is likely caused by the best-fit model being slightly too broad (thus
underestimating the core flux and overestimating the wings), due ultimately to
limitations associated with the set of basis components. This will be addressed
in future data releases by updating the basis components (see
section~\ref{sec:newBases}).

\begin{figure}
\centering
  \includegraphics[width=0.45\textwidth]{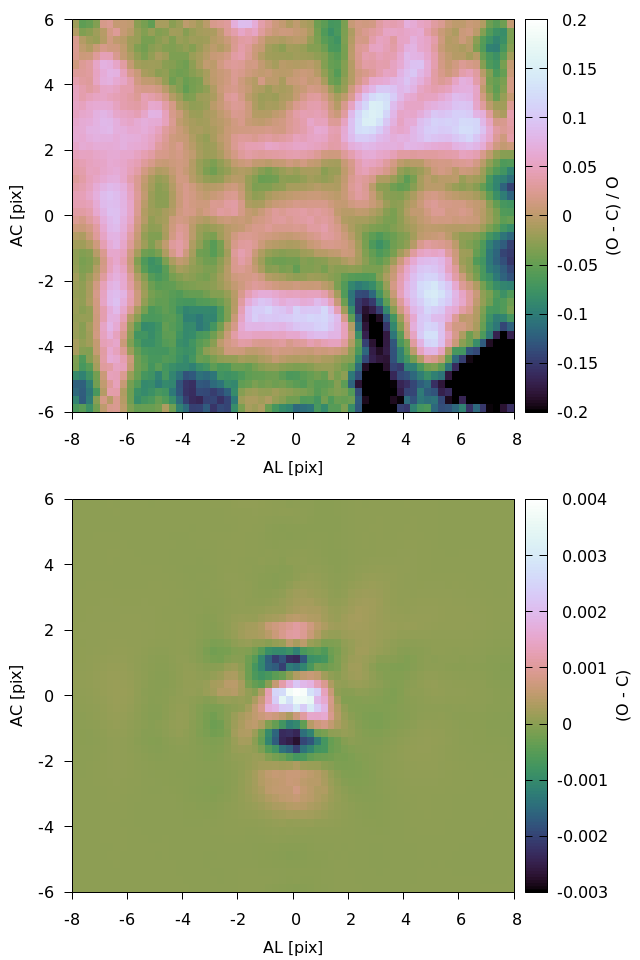}
  \caption{
  Median relative (top) and absolute (bottom) residuals about the model
  for a selection of 2D observations used in the PSF calibration. Observations
  are selected from one calibration unit over a ten revolution period, and span
  the magnitude range $12 \lesssim G \lesssim 13$.}
    \label{fig:psfResid}
\end{figure}

Figure~\ref{fig:psfResid} shows the analogous plots of the 2D residuals about
the PSF model, for the WC0 calibration of the same device, gate,
FOV, and time range.
In this case the corresponding magnitude range is $12 \lesssim G
\lesssim 13$.
The residuals about the PSF model are much stronger and show
significant structure in the core, with departures up to 20\%. The residuals are
also highly dependent on AC rate, with more significant departures up to
30--40\% at times in the steep parts of the profile. The residuals have weaker
dependence on AC rate for the shorter gates, indicating a problem with the PSF
model regarding the AC rate dependence. This is discussed in detail in
section~\ref{sec:psfAcRateIssue}.

\begin{figure}
\centering
  \includegraphics[width=0.45\textwidth]{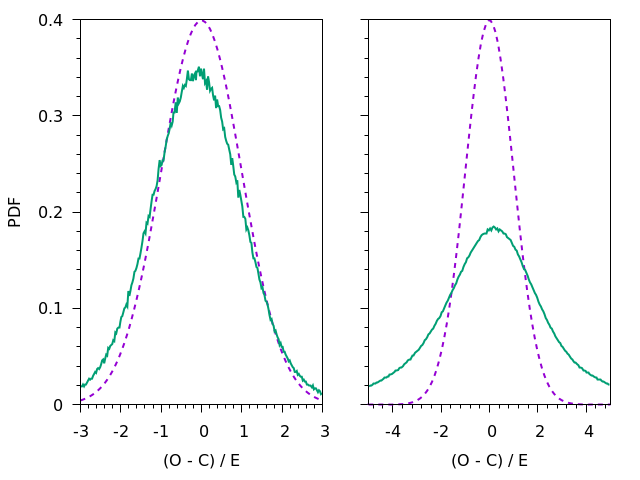}
  \caption{Distribution of the normalised residuals about the model, for
  selected 1D LSF (left) and 2D PSF (right) calibrations. The data is plotted
  in solid green lines, and the dashed purple line shows the unit Gaussian.}
    \label{fig:errorDistributions}
\end{figure}

Finally, in Figure~\ref{fig:errorDistributions} we show the distribution of
normalised residuals about the model (solid green lines), for the LSF (left) and
PSF (right) datasets presented in
Figures~\ref{fig:lsfResid}~and~\ref{fig:psfResid}. The unit Gaussian is plotted
with a dashed purple line.
These indicate that for neither dataset are the departures from the model
consistent with the estimated errors on the observations, indicating that the
model is incomplete and/or the errors on the observations are not correctly
estimated. The inconsistency is much stronger for the PSF, and this is also
reflected in the high value of the $\chi^2/\nu$ statistic shown in
Figure~\ref{fig:gofEvolution}. While to some extent
it is known that the model is not complete (e.g. it includes
only linear effects), the large discrepancy between the LSF and PSF suggests a
significant component of the PSF model may be missing. Again, this is dicussed
in section~\ref{sec:psfAcRateIssue}.

\subsection{Calibration uncertainties}
In addition to the solution for the PLSF parameters the calibration
pipeline computes the associated covariance matrix, by propagating the uncertainties on
the observations through the partial and running solutions. This information can
be used to compute the covariance matrix on the basis component amplitudes at a
particular location in the observation parameter space, and finally to compute
the covariance matrix on the PLSF samples. This provides an estimate of the
calibration uncertainty and can reveal regions of the parameter space that are
less well constrained.

\begin{figure}
\centering
  \includegraphics[width=0.45\textwidth]{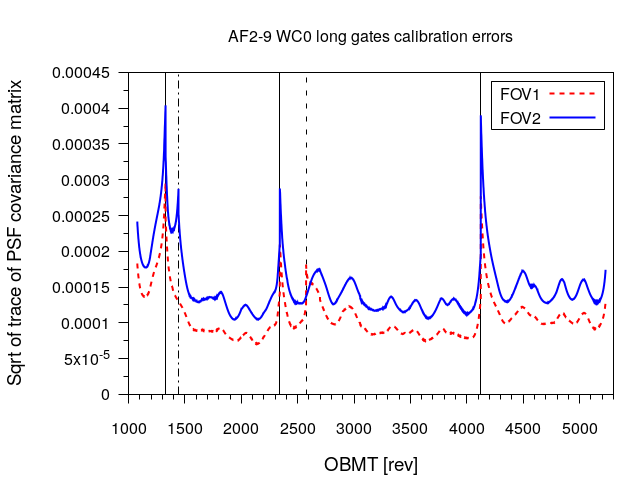}
  \includegraphics[width=0.45\textwidth]{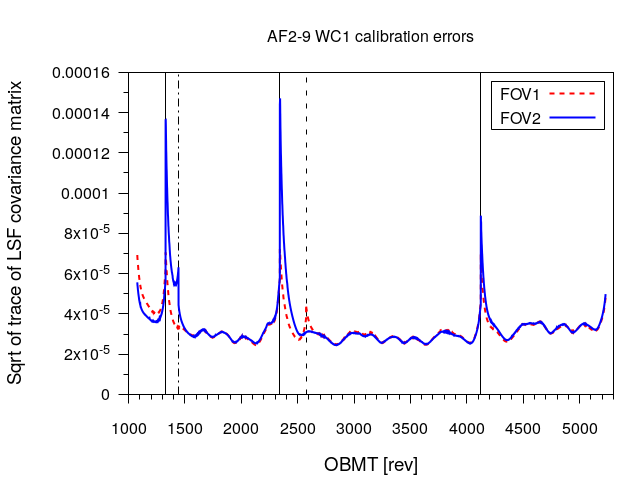}
  \caption{Time evolution in the calibration uncertainty, quantified using
  the square root of the covariance matrix on the PLSF samples at a
  particular point in the observation parameter space. The vertical lines
  indicate resets of the PLSF running solution, as indicated in
  Table~\ref{tab:lsfpsf_resets}.}
    \label{fig:errorEvolution}
\end{figure}

A useful diagnostic is the square root of the trace of the covariance matrix on
the PLSF samples---equivalent to the standard deviation of the sum of the
samples. This quantifies the calibration uncertainty in a single number that is
useful for investigating trends. For example, in Figure~\ref{fig:errorEvolution}
we depict the evolution of this quantity over the whole EDR3 time range, for
both the long gate PSF calibrations (top panel) and LSF calibrations (bottom
panel), averaged over all AF2-9 devices and split by FOV.
The calibration uncertainty shows signficant time-varying features. It is
largest immediately before or after resets of the PLSF running solution; this
is because at these times the running solution is less well constrained, having
been produced from the merger of fewer partial solutions. In the periods between
the solution resets there are significant peaks and troughs. The troughs
coincide with Galactic plane scans, when the rate of observations is much higher
and covers a wider range of $\nu_\text{eff}$ leading to improved constraint
on the model parameters.
For the LSF model both FOVs achieve a similar level of constraint, however for
the PSF model FOV2 is systematically less well constrained than FOV1. This is
likely due to the fact that the PSF model uses different sets of basis
components to model FOV1 and FOV2; the two sets are unlikely to offer the same
level of accuracy in reproducing the observations.

\begin{figure}
\centering
  \includegraphics[width=0.45\textwidth]{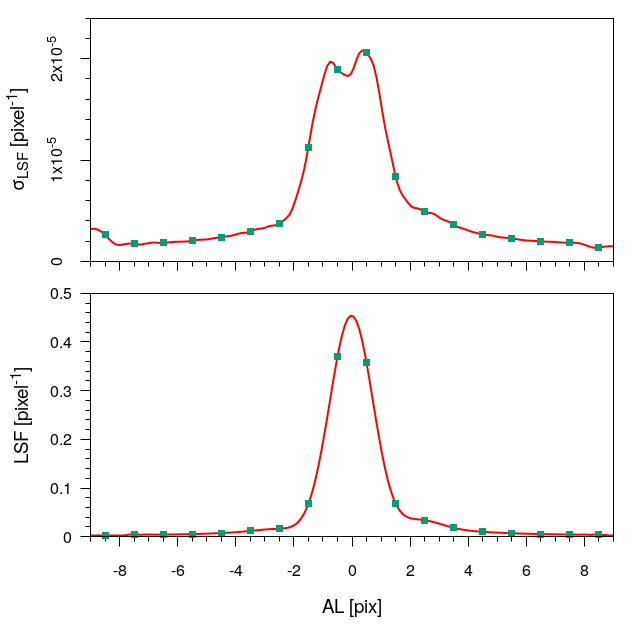}
  \caption{AL variation in the calibration errors for a single  
  representative realisation of the LSF model. The top panel depicts the
  standard deviation of the LSF model from propagation of the covariance on the
  LSF parameters; the lower panel depicts the corresponding LSF value. There is
  notable structure in the top panel, as explained in the text. The green
  points correspond to the locations of the samples for which the full
  covariance matrix is shown in Figure~\ref{fig:lsfCovar}.}
    \label{fig:lsfErrors}
\end{figure}

Figure~\ref{fig:lsfErrors} depicts the AL variation in the calibration
uncertainties for a single realisation of the LSF model. In the top panel the
red line indicates the standard deviation of the LSF model shown in the bottom
panel. The green points indicate the locations of the 18 samples that would be
observed in a real \textit{Gaia} window. There is some notable structure present
here. First, the dip in the core is due to the way that location errors on the
observations used in the LSF calibration are propagated to the errors on the
sample values; this inflates the errors in the steep parts of the profile, such
that the steep wings are less well constrained than the flatter core. Second,
the increase in $\sigma_{\text{LSF}}$ around $\text{AL}\sim-8.5$ arises because
the observations used to calibrate this LSF (which corresponds to FOV1, CCD row
4, strip AF5) are systematically shifted from the window centre such that the
LSF is this region is not well sampled by the data and the model is less well
constrained. This is a general feature of the way that \textit{Gaia} operates;
the FOV2 observations in the same device are offset in the opposite direction
and the corresponding LSF model is well constrained in this region (but not at
$\text{AL}\sim8.5$).

\begin{figure}
\centering
  \includegraphics[width=0.45\textwidth]{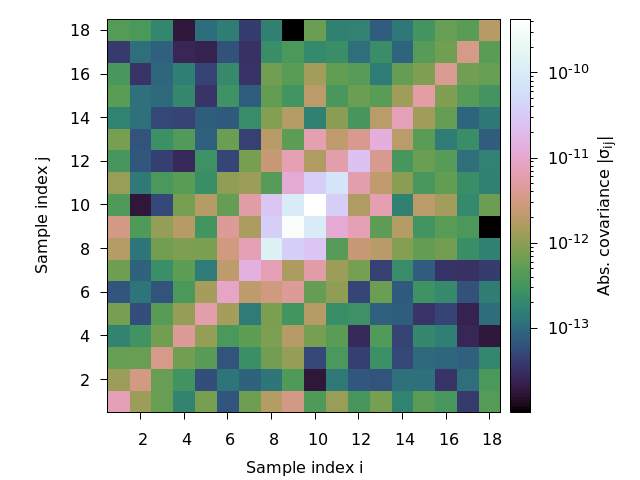}
  \caption{Covariance matrix for the model LSF samples shown in
  Figure~\ref{fig:lsfErrors}, in terms of the absolute value of the covariance
  for clarity of plotting. There are significant non-zero covariance terms
  between pairs of model samples, particularly in the core of the profile. This
  is in stark constrast to the observations used to calibrate the model, for
  which the errors on each sample are independent.}
    \label{fig:lsfCovar}
\end{figure}

\begin{figure}
\centering
  \includegraphics[width=0.45\textwidth]{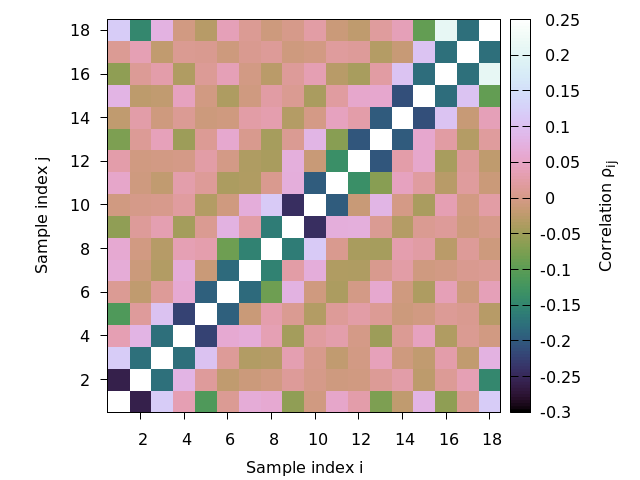}
  \caption{Correlation statistic $\rho_{ij} =
  \sigma_{ij}/(\sigma_i \sigma_j)$ for the model LSF sample uncertainties shown in
  Figure~\ref{fig:lsfErrors}. The diagonal elements all have $\rho_{ij} = 1$ and
  have been eliminated in order to avoid stretching the colour scale. The main
  feature is a negative correlation on the uncertainties between neighbouring
  samples.}
    \label{fig:lsfCorr}
\end{figure}

Finally, Figure~\ref{fig:lsfCovar} depicts the full covariance matrix for the
model samples indicated by the green points in Figure~\ref{fig:lsfErrors}. We
note the presence of large off-diagonal terms, particularly in the LSF core
(central part of the plot), that indicate the existence of significant
covariances between the model samples. While for a given observation the noise
on each sample is independent, the same is not true of the samples drawn from
the calibrated PLSF model. This is a statistical property of the model that
may need to be considered by downstream \textit{Gaia} systems that make use of
the PLSF calibration.
Figure~\ref{fig:lsfCorr} provides a complementary plot of the correlation
statistic, which gives a better impression of the relative importance of the
covariance terms and distinguishes between negative and positive values. The
most important feature is the presence of a negative correlation on the
uncertainties between neighbouring samples.

\section{Discussion}
The PLSF modelling and calibration performed for \textit{Gaia} EDR3
represents a major step forwards relative to DR1 and DR2, in particular with the
activation of the time and colour dependences, introduction of a full 2D PSF
model, and the first closure of the iterative loop with the astrometric
solution.
However, the PLSF models as currently formulated are known to be incomplete,
excluding, for example, magnitude dependent effects, and being subject to
various compromises and approximations that are the consequence of unavoidable
limitations in the data processing chain.
%
%
%
%
Extensive analysis of the calibration results carried out since the pipeline
was executed has clarified the directions for future development, and also
revealed some defects with the current modelling, alluded to earlier in this
paper, that will be corrected in later data releases.
In this section we discuss some of these issues.
\subsection{Missing AL rate dependence in PSF model \label{sec:psfAcRateIssue}}
Undoutedly the most significant problem with the current PLSF modelling is
the presence of a systematic error in the PSF model, particularly for long gates
(gate 11, 12 and 0), that has a strong correlation with AC rate. This manifests
as a bimodality in the calibrated PSF that is evident in
Figure~\ref{fig:acRateVariation} and also reflected in the spatial structure in
the residuals (Figure~\ref{fig:psfResid}) and in their distribution
(Figure~\ref{fig:errorDistributions}, right panel).
This issue has undergone extensive investigation since the calibration was
performed, and is now understood to be caused largely by the presence of an
additional PLSF dependence that is absent from the model, that of the
along-scan rate.

%
In the TDI mode in which \textit{Gaia} operates, the (fixed) rate of parallel
charge transfer must be closely matched to the rate at which stellar images
drift across the CCDs, which in turn is determined by a combination of the scan
rate and the AL angular pixel scale.
Any mismatch between the two has the effect of broadening the apparent PSF in
the AL direction, as the stellar image gradually lags behind or moves ahead of
the integrating charge during exposure.
The AL scan rate is continuously adjusted to match the TDI rate as closely as
possible, and it is subject to both systematic and random variations induced by
the \textit{Gaia} scan law \citep[see][section 5.2]{2016A&A...595A...1G} and
various disturbances such as micro-meteorites and thermo-mechanical `clanks'.
The component induced purely by the scan law varies sinusoidally with a period
of one revolution and an amplitude of up to \textasciitilde0.03 pixels per
second.
It is caused by a rotation of the field arising from the precessional motion and
not, for example, by a variation in the satellite rotation rate. The same effect
gives rise to the AC rate modulation, which is much larger. The scan-law
component of the AL rate varies in strength and sign across the focal plane, is
out of phase by $\pi/2$ with the associated AC rate modulation in the same
telescope and is out of phase between the two telescopes by $106.5^{\circ}$.
The AL angular pixel scale also differs systematically between the two
telescopes due to focal length differences, and varies across the focal plane
depending on the distance from the optical axis.
The result is that the AL image rate differs significantly from the parallel
charge transfer rate, which violates the assumptions of both the LSF and PSF
model and leads to some importance consequences.

\begin{figure*}
\centering
  \includegraphics[width=0.95\textwidth]{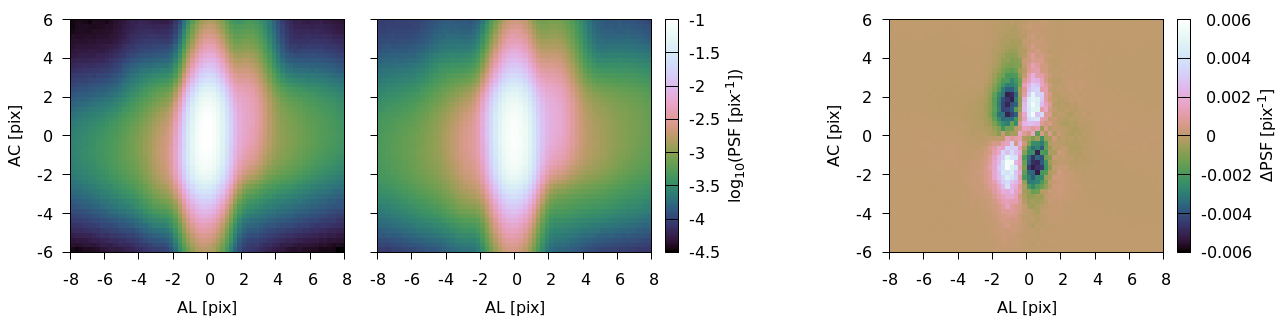}
  \caption{
  Effects of AC and AL rate variations on the observed PSF. The two panels on
  the left depict many stacked observations from FOV1 ROW1 AF6 gate 0, for two
  narrow ranges of AC rate at high negative ($\text{-}1.0 \rightarrow
  \text{-}0.95$ pix/s; left) and high positive ($0.95 \rightarrow 1.0$ pix/s;
  middle) values. These reveal an apparent shear between the two PSFs that is
  clear in a plot of the difference (right panel). This effect is induced by
  non-zero AL rate, as explained in the text.}
    \label{fig:alRate}
\end{figure*}

The observable effect on the LSF is relatively minor, and leads to a slight
broadening of the profile by up to \textasciitilde0.1 pixels that varies over a
revolution; the time dependence in the LSF solution lacks the resolution to
capture this so that in effect the LSF solution fits the average profile, and
the effects on the astrometry and photometry are minimal.
However, the effect on the PSF is stark. The combination of the out-of-phase AL
and AC rates induces a shear on the PSF that varies in sign and magnitude over
the course of one revolution. This is depicted in Figure~\ref{fig:alRate}, which
compares the observed FOV1 PSF (from aligned and stacked observations) at
maximum negative (left panel) and positive (middle panel) values of the AC rate,
half a revolution apart. At the AC rate extrema the AL rate variation is
minimal (due to the $\pi/2$ phase difference), so that only the
zeropoint offset remains and the AL rate takes on a single value. In such a
situation the existing PSF model is in principle able to reproduce the
observations via the AC rate dependence, and it is interesting to note that the
difference between the two plots (shown in the third panel) closely matches one
of the low-order FOV1 basis components shown in
Figure~\ref{fig:genshapelets_0_3}, indicating that the model is attempting to
fit the average shearing. However, in general the relation between the AL and AC
rate is not monotonic, and no model that considers only the AC rate can
accurately reproduce the PSF.
This effect is much stronger for longer integration times as this leads to
larger total AL and AC displacements. In terms of the CCD gates, the effect is
strongest for gate 0 and negligible for gate 10, so the impact on the astrometry
and photometry is expected to be largest for sources in the corresponding
magnitude range of $11.5 \lesssim G \lesssim 13$.

The combined effects of AL and AC rate could be naively incorporated into the
PSF model by adding the AL rate as another dimension in the observation
parameter space. However, this approach faces numerous problems due to the curse
of dimensionality: The sparsity of the observations within the parameter space
makes the model difficult to constrain, and the explosion in the number of PSF
parameters (and covariances) makes the calibration too computationally
demanding.
Fortunately, the effects of the AL and AC rate can be modelled quite
successfully from first principles as a convolution with a top hat kernel of the
appropriate width and orientation. This avoids the need to calibrate the effects
empirically and thus eliminates the associated PSF parameters. An analytical
model that performs this has been developed and integrated into the PLSF
pipeline for use in future \textit{Gaia} data releases. This will be described
in more detail in a dedicated publication.
\subsection{Uncertainty estimation for PSF observations}
The systematic error in the calibrated PSF model was also found to be partly due
to inappropriate weighting of the samples used to constrain the model. As
described in section~\ref{sec:obs_prep}, the estimated error on each sample
includes a contribution to account for uncertainty on the predicted location of
the source. This increases the error in steeper parts of the profile where the
effects of location errors are more significant.
The motivation is to down-weight observations with larger location uncertainty,
which include a greater fraction of outliers such as undetected close pairs.
However, this also has the undesirable effect of reducing the constraint in the
steep parts of the profile for all sources, and results in a solution that
preferentially fits the flatter regions of the profile that do not carry as much
astrometric constraint (although these regions are more useful to the
photometry). This can be seen clearly in Figure~\ref{fig:lsfErrors}. The effect
is much greater on the PSF calibration than the LSF because for the PSF the
uncertainty on the AC location is also included, and in the 2D observations
there are relatively fewer samples in the steep parts of the profile than for
the 1D observations so they are further down-weighted.
For future \textit{Gaia} data releases this procedure will be avoided, and
outlying observations will be handled by other methods.
\subsection{Incomplete colour calibration for PSF}
Both the astrometric and photometric processing encountered issues that suggest
the colour calibration of the PSF was less successful than for the LSF. In the
astrometric processing, this manifested as a strong residual chromaticity for
the 2D observations compared with 1D \citep[see][appendix A.4]{EDR3-DPACP-128}.
In the photometric processing, a colour term was present in the $G$ band for
$G<13$ (corresponding to 2D observations) but not for fainter, 1D observations
\citep[see][section 7]{EDR3-DPACP-117}. The reason for this is most likely due
to differences in the basis components used by the LSF and PSF models and the
major dependences that are active. The LSF model used 25 1D basis components
that are tuned to empirically model only the colour dependence, whereas the PSF
model used 30 basis components that are tuned to model both the colour and AC
rate dependences. The AC rate dependence has a very large effect on the profile,
which for a fixed number of basis components in effect reduces the ability to
model colour variations. This issue will hopefully be solved in later data
releases by the analytic modelling of the AC rate (and AL rate) dependence in
the PSF, which means that the 2D basis components can be tuned for empirical
modelling of the colour dependence only.
\subsection{Use of a reference PLSF calibration \label{sec:refLsfPsf}}
A fixed reference PLSF calibration is used at several points in the
calibration pipeline---see section~\ref{sec:obs_prep}. The reference
calibration was solved in a separate procedure by fitting the PLSF model to
observations selected from a stable period of low contamination, using an
iterative solution to reject outliers and update the normalisation of each
observation.
This produces a clean calibration that is suitable for use in the main pipeline
for outlier detection, computing corrections to the observation errors, and in
normalising the observations to account for flux falling outside the window
area.
The last of these applications contains some subtleties that require further
discussion.

The LSF and PSF models are normalised to unity, over the whole AL dimension for
the LSF and over the AL and AC dimensions for the PSF. Due to the finite extent
of the windows, each observation sees only a fraction of the total flux received
from a source---the enclosed energy fraction. The observations that are used to
calibrate the LSF and PSF models must be normalised to match the requirements of
the model, and if the enclosed energy fraction is not accurately accounted for
then the model can end up trying to force flux into or out of the unobserved
regions beyond the window boundary in order to better fit the observed samples,
which invariably introduces artefacts in the shape.

One way to estimate the enclosed energy fraction for a particular observation of
a known $G$ band magnitude source is to invert the photometric calibration and
compute the associated instrumental flux. This provides an estimate of the total
number of photoelectrons in the detector, and the ratio between the observed
number of photoelectrons within the window and the total number gives the
enclosed energy fraction. We note that for the LSF the fraction of flux lying
outside of the window in the AC direction (the `AC flux loss') must also be
accounted for---this is included as part of the photometric calibration
\citep{EDR3-DPACP-117}.
This is conceptually similar to the closure of the iterative loop with the
astrometric calibration, where the use of predicted locations to calibrate the
PLSF enables a greater consistency and separation of effects between the two
systems--though care must be taken to avoid circularity and feedback of
systematic errors.

However, this method cannot be used, owing mainly to the fact that the PLSF
calibration pipeline runs before the photometric calibration in the
\textit{Gaia} processing chain.
A workaround that has been adopted for EDR3 is to use the fixed reference
PLSF calibration to estimate a physically reasonable value for the enclosed
energy fraction of each observation, from the sum of the model samples within
the window area.
This value is accurate to around the \textasciitilde1\% level, which is
sufficient to allow the PLSF models to converge to a solution that
accurately reproduces the PLSF shape. This leaves systematic errors in the
PLSF normalisation of around \textasciitilde1\%, which propagate to the
estimated fluxes of observations derived from PLSF fitting, and are then
corrected as part of the photometric calibration. So this method has limited
impact on both the astrometry and photometry and provides a suitable solution
for EDR3.

However, certain improvements in the CCD reductions and instrument modelling
that are under development for future data releases cannot tolerate systematic
errors in the PLSF calibration at this level. These include, for example,
the modelling of non-linear magnitude-dependent effects that are
highly sensitive to the number of photoelectrons in each pixel.
%
%
Therefore, future developments to the PLSF calibration will aim to reduce
these systematic errors further. This may be done by introducing limited
calibrations of the system throughput earlier in the pipeline, in order to
perform a rudimentary inversion of the photometric calibration that will allow
these effects to be properly calibrated in the PLSF and not enforced
\textit{a priori}.
\subsection{Tracking of rapidly evolving instrument state
\label{sec:discRunningSolutionAfterDecontamination}}
The running solution that is used to calibrate the time dependence of the
PLSF parameters (see sections~\ref{sec:timedep} and~\ref{sec:runningSol})
successfully tracks the gradually evolving instrument state during nominal
periods. However, shortly after a decontamination there is a period of thermal
instability during which the instrument state evolves rapidly and the running
solution fails to converge on the instantaneous solution (see
Figure~\ref{fig:fwhm_decont5}). This situation can be improved either by
reducing the time decay constant in the running solution, to allow faster
convergence at the expense of reduced smoothing of noise, or by excluding a
greater segment of the data during the period of thermal instability. This will
be investigated for future \textit{Gaia} data releases.
\subsection{Use of a single colour parameter}
The effective wavenumber $\nu_{\text{eff}}$ is a good parameterisation of the
source colour as the chromatic shifts for sources with normal stellar SEDs are
expected to be linear in $\nu_{\text{eff}}$ \citep{LL:JDB-028}. It is also
convenient from the calibration point of view, as it is a single number and its
value is restricted to a relatively narrow range.
However, it does introduce limitations when processing sources that have
non-stellar SEDs, such as quasars or emission line objects. Other extended
parameterisations of the source colour, such as the spectral shape coefficients
\citep[see][section 4.4]{EDR3-DPACP-117} may offer better colour
parameterisations for these sources. However, this complicates the calibration
as such objects are rare, and a degradation of the modelling in these cases may
have to be accepted.
\subsection{Improvements for DR4 and beyond \label{sec:c04}}
In addition to dealing with the various problems described earlier in
this section, there are a range of improvements to the PLSF models that are
planned for future data releases. A brief description of these is presented in
this section. At the time of writing, some have already been implemented for DR4
and are in the final stages of testing. These will be described in detail in a
dedicated publication, but in anticipation of these advances they are
summarised here.
\subsubsection{Analytic modelling of AL and AC rate effects}
The effects of AL and AC rate present significant problems to the EDR3 PSF
modelling in particular, and cannot be calibrated empirically (see
section~\ref{sec:psfAcRateIssue}). However, unlike the other dependences the
effects of AL and AC rate on the PLSF can be derived from first principles,
and modelled accurately as a convolution with a top hat of an appropriate width
and orientation. This will be implemented in future PLSF modelling and will
lead to improved PLSF reconstruction with reduced number of parameters.
\subsubsection{Calibration of AL and AC shift parameters \label{sec:alshift}}
The LSF model presented in Equation~\ref{eq:lsf} includes a parameter $u_0$
that corresponds to a pure shift of the LSF profile in the AL direction
without a change in shape. The PSF model has an additional parameter
corresponding to shifts in the AC direction.
The LSF model is non-linear in $u_0$ which makes the calibration of it
significantly more challenging, and in EDR3 this parameter is not calibrated and
instead is fixed at zero. This has the effect of allowing any shifts in the
profile to be absorbed into the calibration of the other parameters and modelled
by appropriate weighting of the $H_n$ basis components.

The PLSF model is calibrated to the predicted locations of sources provided
by the source astrometry combined with the attitude and geometric instrument
calibrations.
In principle a perturbation in the geometric calibration leads to a displacement
of the predicted locations of sources that is compensated for by a
similar shift in the PLSF calibration. There is thus a degeneracy between the
geometric instrument calibration and the PLSF origin. The key to breaking
this degeneracy is to make an accurate calibration of the $u_0$ parameter,
including its variation with time, source colour, and other parameters. This can
then be used to enforce constraints on the PLSF model that allow a full
separation of the PLSF and geometric calibrations. This will be implemented
for future data releases.
\subsubsection{Inclusion of magnitude-dependent effects \label{sec:magDep}}
The PLSF model includes only the linear component of the CCD
response arising purely from optical effects. However, there are numerous
second-order components of the CCD response that are non-linear in the source
flux, and that have complex dependences on both the flux and other parameters,
resulting in redistribution and/or loss of charge entirely from the window.
These effects are introduced by several phenomena that manifest in the
\textit{Gaia} CCDs, including deflection of incoming photoelectrons by large
charge packets (the `brighter-fatter' effect, \citet{2014JInst...9C3048A}),
blooming (in conjunction with strong spatial variations in the performance of
the anti-blooming drains), and AL variation in the pixel well capacity (which in
conjunction with the TDI mode of operation introduces non-linearity to the CCD
response at high signal levels). The CCDs are also affected by CTI in the image
section, which redistributes charge in the AL direction, and CTI in the serial
register, which redistributes charge in the AC direction.
CTI in particular is an important component that has complex dependences on the
source magnitude, CCD illumination history and, for serial CTI, the readout
sequence, which varies from one TDI line to the next and is determined by the
distribution of windows along the serial register. The existence of a
supplementary buried channel in the CCD image section is expected to introduce a
break in CTI behaviour for faint stars.
Extensive work was done pre-launch on quantifying the effects of CTI on
observations of stars \citep[e.g.][]{2011MNRAS.414.2215P}, and continued
investigation using in-flight data will benefit greatly from an accurate
calibration of the linear part of the PLSF.

In terms of modelling all of these effects, it is clear that simply extending
the PLSF parameterisation to include source flux is not sufficient. Instead, it
is anticipated that the PLSF model will continue to include only the linear
part of the CCD response, and the various non-linear components will be
incorporated via a separate forward-model of the CCD pixel-level behaviour.
In this scenario, the existing PLSF model will predict the spatial
distribution of incoming photons, which in turn provides the inputs for the next
stage of modelling the CCD response.
\subsubsection{Improvements to the basis components \label{sec:newBases}}
The 1D basis components used to model the LSF (and, indirectly, the PSF) for
EDR3 were produced pre-launch using simulations of the optical system,
as described in section~\ref{sect:lsf} and in detail in \citet{LL:LL-084}
(see also Figures~\ref{fig:lsfbases_0_3} and~\ref{fig:genshapelets_0_3}).
While these have been sufficient for EDR3, for future data releases we are
investigating whether updates to the basis components may offer significant
improvements to the PLSF reconstruction.
Recent analysis has revealed some minor numerical artefacts in the optical model
discretisation and the interpolation of the individual bases, as explained in
\citet{LL:PMN-012}. The bases are also not fully orthogonal, and are computed in
the AL direction only.
More significant is the possibility that the optical model fails to include some
important elements of the instrument. This can result in basis components that
fail to reproduce certain variations present in the data, leading to systematic
errors.
For example, it is now known that the wavefront error includes a significant
systematic component arising from mirror polishing artefacts in the primary
mirrors, and this is absent from the original optical model.
We therefore intend to revise the simulation code to fix these various issues,
and to investigate the limitations of the basis components to identify any
elements missing from the optical model.
\subsubsection{Independent calibration of SM}
The calibration of the PSF for the SM instrument faces numerous challenges, as
described in section~\ref{sec:sol_copying}, and for EDR3 an independent
calibration of SM was not possible. While in EDR3 this has limited impact, data
products planned for future releases may wish to make greater use of the SM
observations and will require an accurate calibration of the PSF.
For example, SM observations are not currently used in either the
astrometric or photometric solutions for each source. Also, certain selected
dense regions of the sky are scanned using a special mode where, in addition to
the normal source detection and windowing, the full SM images are downlinked
without windowing. These data are presently not used, and any future processing
of them will require a dedicated calibration of the SM PSF.
Future developments of the PSF model and pipeline will aim to overcome the
challenges presented by SM so that an independent calibration can be achieved.
\subsubsection{Bootstrapping of attitude and geometric calibration}
One of the challenges faced in the production of the PLSF calibration for
EDR3 was the need to obtain predicted locations for sources during the first
iteration of the solution, which occurred before AGIS had computed the required
attitude and geometric calibrations.
This had to be overcome by using the calibrations from DR2 as the starting
point, supplemented with calibrations taken from the \textit{Gaia} realtime
pipeline to cover the additional time segment after the end of DR2 and before
the end of EDR3. This introduced an inhomogeneity in the data that resulted in
some systematic errors in the first iteration of the PLSF calibration that
required an additional iteration with AGIS to reduce.

This reliance on inputs from previous cycles and other sources, computed using
different generations of the instrument models, to initialise the PLSF
calibration risks introducing artefacts and systematic errors that may be hard
to eradicate. For future data releases, a new bootstrapping of the attitude and
geometric calibrations is being developed that will provide a homogeneous and
consistent set of inputs for initialising the PLSF calibration.
\subsubsection{Far PSF calibration}
The PLSF models implemented for EDR3 cover only the core of the
profile contained within the \textit{Gaia} windows for nominal observations.
There are several applications within the \textit{Gaia} data processing systems
that require a calibration of the PSF over a much wider range. These include the
analysis of very bright stars for which the entire core region is saturated, and
the background modelling for normal stars in the vicinity of bright stars.
For various reasons the PSF model presented in this paper cannot be easily
adapted to model the extended profile of bright stars. For future data
processing cycles a new model is under development that aims to provide a
suitable calibration of the extended PSF.
\section{Conclusions}
The PLSF modelling and calibration carried out for \textit{Gaia} EDR3
represents a major step forwards in the data processing, and will
contribute to reduced systematic errors in the core mission data
products.
This is reflected in both the astrometric and photometric solutions for EDR3,
which see improvements relative to DR2 beyond those expected from the increased
number of observations alone (see e.g.~section 5.4 and appendix A.1 in
\citet{EDR3-DPACP-128}, and section 9.5 in \citet{EDR3-DPACP-117}).

In this paper we have presented a detailed description of the models, the
pipeline and the calibration products that is necessary for a complete
understanding of the EDR3 contents and survey properties.
These developments are part of an ongoing process of gradual refinement and
improvement as the instrument modelling increases in fidelity and we gain a
deeper understanding of the data. Further significant improvements are expected
in the future data releases.

\begin{acknowledgements}
The authors would like to thank the referee Jay Anderson for insightful comments
that have improved the quality and clarity of this paper.

This work has made use of data from the European Space Agency (ESA) mission
{\it Gaia} (\url{https://www.cosmos.esa.int/gaia}), processed by the {\it Gaia}
Data Processing and Analysis Consortium (DPAC,
\url{https://www.cosmos.esa.int/web/gaia/dpac/consortium}). Funding for the DPAC
has been provided by national institutions, in particular the institutions
participating in the {\it Gaia} Multilateral Agreement. The \textit{Gaia}\
mission website is:
\url{http://www.cosmos.esa.int/gaia}.

The work described in this paper has been financially supported by
the United Kingdom Particle Physics and Astronomy Research Council (PPARC), the
United Kingdom Science and Technology Facilities Council (STFC), and the
United Kingdom Space Agency (UKSA) through the following grants to the
University of Bristol, the University of Cambridge, the University of
Edinburgh, the University of Leicester, the Mullard Space Sciences Laboratory
of University College London, and the United Kingdom Rutherford Appleton
Laboratory (RAL): PP/D006511/1, PP/D006546/1, PP/D006570/1, ST/I000852/1,
ST/J005045/1, ST/K00056X/1, ST/K000209/1, ST/K000756/1, ST/L006561/1,
ST/N000595/1, ST/N000641/1, ST/N000978/1, ST/N001117/1, ST/S000089/1,
ST/S000976/1, ST/S001123/1, ST/S001948/1, ST/S002103/1, and ST/V000969/1;
%
the Spanish Ministry of Economy (MINECO/FEDER, UE) through grants
ESP2016-80079-C2-1-R, RTI2018-095076-B-C21 and the Institute of Cosmos Sciences
University of Barcelona (ICCUB, Unidad de Excelencia ’Mar\'{\i}a de Maeztu’)
through grants MDM-2014-0369 and CEX2019-000918-M;
the German Aerospace Agency (Deutsches Zentrum für Luft- und Raumfahrt
e.V., DLR) through grants 50QG0501, 50QG0601, 50QG0602, 50QG0701, 50QG0901,
50QG1001, 50QG1101, 50QG1401, 50QG1402, 50QG1403, and 50QG1404 and the Centre
for Information Services and High Performance Computing (ZIH) at the Technische
Universität (TU) Dresden for generous allocations of computer time;
the Agenzia Spaziale Italiana (ASI) through contracts I/037/08/0,
I/058/10/0, 2014-025-R.0, 2014-025-R.1.2015 and 2018-24-HH.0 to the Italian
Istituto Nazionale di Astrofisica;
and the Swedish National Space Board (SNSB/Rymdstyrelsen).
%
AB additionally acknowledges financial support from the Netherlands Research
School for Astronomy (NOVA).
Finally, the authors also acknowledge the computer resources from MareNostrum,
and the technical expertise and assistance provided by the Red Espa\~{n}ola de
Supercomputaci\'{o}n at the Barcelona Supercomputing Center, Centro Nacional de
Supercomputaci\'{o}n.

\end{acknowledgements}

\bibliographystyle{aa}
\bibliography{refs}

\end{document}